\documentclass[12pt,preprint]{aastex}
\usepackage{longtable}
\usepackage{color}
\everymath{\rm}
\everydisplay{\sf}

\begin{document}

\title{Abundances of Galactic Anticenter Planetary Nebulae and the Oxygen Abundance Gradient in the Galactic Disk\altaffilmark{1}}

\author{R.B.C. Henry\altaffilmark{2,3}}
\affil{H.L. Dodge Department of Physics \& Astronomy, University of Oklahoma, Norman, OK 73019, USA}
\email{henry@nhn.ou.edu}

\author{Karen B. Kwitter\altaffilmark{3}}
\affil{Department of Astronomy, Williams College, Williamstown, MA 01267, USA}
\email{kkwitter@williams.edu}

\author{Anne E. Jaskot\altaffilmark{4}} 
\affil{Department of Astronomy, Williams College, Williamstown, MA 01267, USA}
\email{ajaskot@umich.edu}

\author{Bruce Balick}
\affil{Department of Astronomy, University of Washington, Seattle, WA 98195, USA}
\email{balick@astro.washington.edu}

\author{Michael A. Morrison}
\affil{H.L. Dodge Department of Physics \& Astronomy, University of Oklahoma, Norman, OK 73019, USA}
\email{morrison@nhn.ou.edu.edu}

\and

\author{Jacquelynne B. Milingo\altaffilmark{2,3}}
\affil{Department of Physics, Gettysburg College, Gettysburg, PA 17325, USA}
\email{jmilingo@gettysburg.edu}

\altaffiltext{1}{Partially based on observations obtained with the Apache Point Observatory 3.5-meter telescope, which is owned and operated by the Astrophysical Research Consortium.}
\altaffiltext{2}{Visiting Astronomer, Cerro Tololo Interamerican Observatory, National
Optical Astronomy Observatory, which is operated by the Association of
Universities for Research in Astronomy (AURA), Inc., under cooperative
agreement with the National Science Foundation.}
\altaffiltext{3}{Visiting Astronomer, Kitt Peak National Observatory, National Optical
Astronomy Observatory, which is operated by the Association of Universities
for Research in Astronomy (AURA), Inc., under cooperative agreement with the
National Science Foundation.}
\altaffiltext{4}{Now at Department of Astronomy \& Astrophysics, University of Michigan, Ann Arbor, MI 48109, USA}

\begin{abstract}

We have obtained spectrophotometric observations of 41 anticenter planetary nebulae (PNe) located in the disk of the Milky Way. Electron temperatures and densities, as well as chemical abundances for He, N, O, Ne, S, Cl, and Ar were determined. Incorporating these results into our existing database of PN abundances yielded a sample of 124 well-observed objects with homogeneously-determined abundances extending from 0.9-21 kpc in galactocentric distance. We performed a detailed regression analysis which accounted for uncertainties in both oxygen abundances and radial distances in order to establish the metallicity gradient across the disk to be: $12+log(O/H)=(9.09\pm.05) - (0.058\pm.006) \times R_g$, with $R_g$ in kpc.  While we see some evidence that the gradient steepens at large galactocentric distances, more objects toward the anticenter need to be observed in order to confidently establish the true form of the metallicity gradient. We find no compelling evidence  that the gradient differs between Peimbert Types I and II, nor is oxygen abundance related to the vertical distance from the galactic plane. Our gradient agrees well with analogous results for H~II regions but is steeper than the one recently published by \citet{stanghellini10} over a similar range in galactocentric distance. A second analysis using PN distances from a different source implied a flatter gradient, and we suggest that we have reached a confusion limit which can only be resolved with greatly improved distance measurements and an understanding of the natural scatter in oxygen abundances. Finally, a consideration of recently published chemical evolution models of the Galactic disk suggests that reconciling the current range in published oxygen gradients is necessary for adequately constraining parameters such as the surface density threshold for star formation and the characteristic timescale for disk formation.
\end{abstract}

\keywords{ISM: abundances, abundances, planetary nebulae: general, stars: evolution, galaxies: evolution}

\section{Introduction}

The current metallicity of the ISM at any point in the disk of a galaxy serves as an endpoint indicator of all the element synthesis processes that have ever occurred in material currently at that location up to the present. Therefore, computing a successful one-zone chemical evolution model of an arbitrary location using the observed metallicity as a constraint helps us to accurately reconstruct the element enrichment history as well as infer the stellar rates of element production associated with the region.

A chemical evolution model of the entire disk is essentially a linked set of one-zone models organized in order of increasing galactocentric distance, while one of the major constraints on such a model is the observed behavior of metallicity with radial distance, a behavior which is usually found to be continuous, monotonic, arguably linear, and negative., i.e., the abundance gradient. Thus models which successfully predict the abundance gradient help us understand: (1) the local processes referred to above plus the distribution of matter across the disk which fuels them; and (2) the metallicity and rate of infall of halo material which formed the disk at each point out from the nucleus. A general review of galactic abundance gradients in disk galaxies and ellipticals can be found in \citet{henry99}, while many of the papers cited below serve as valuable updates to this article.

Abundance gradients in the Milky Way disk are commonly tracked by measuring the abundance of metals such as O, Ne, S, Ar, and Fe as a function of galactocentric distance. The first four elements are most readily (but not exclusively) observed in emission line objects such as H II regions \citep{deharveng00, rudolph06} and planetary nebulae \citep{hkb04, perinotto06, maciel08,stanghellini10}. Oxygen, the focus of this paper, can also be usefully measured in B dwarf stars \citep{smartt97}, while O and Fe can be measured in Cepheid variable stars \citep{andrievsky04}. Most researchers find that the oxygen abundance gradient is linear in the log-normal plane of O versus galactocentric distance and generally ranges between -0.01 to -0.07 dex~kpc$^{-1}$, although a few authors, e.g., \citet{vilchez96, maciel08}, claim that the gradient may flatten out at large radial distances from the Galactic center. Other alpha elements such as Ne, S, and Ar follow similar trends \citep{hkb04, maciel06}.

Planetary nebulae (PNe) are doubly useful probes of chemical composition, as they provide information about low- and intermediate-mass stellar nucleosynthesis affecting He, C, N, as well as archiving abundances of heavier elements like O, Ne, Ar, Cl, and S at the time when the progenitor star formed\footnote{There is some evidence that the original progenitor levels of O and Ne can be altered as a result of CNO processing \citep{pequignot00, wang08, mkhs}. However,  because this idea has yet to be firmly established, in this paper we shall assume that PN oxygen levels are unchanged from the levels inherited by the progenitor star from the ISM at birth.}. Here we assume that the progenitor abundance is equivalent to the ISM value at the time of star formation. For this second group of elements, PN abundances measured at various galactocentric distances have been used frequently to determine the abundance gradient in the disks of the Milky Way and other spiral galaxies \citep{hkb04,maciel06,magrini07,bresolin10,stanghellini10}. 

Recently, PN abundances have been used by Maciel and collaborators [see  \citet{maciel08} and references cited therein for a complete summary] to study the {\it evolution} of the gradient of oxygen and other elements. This particular application follows from the idea that since progenitor stars range in mass between 0.08 and 8~M$_{\odot}$ and therefore collectively probe the interstellar medium at different times over billions of years, abundance and radial distance can now be coupled with age to infer slope changes with time. As a result, the Maciel team claims that the oxygen gradient and the gradients of Ne, S, and Ar have flattened with time. In contrast, however, \citet[S10]{stanghellini10} infer from their study that the gradient in the Milky Way disk has steepened with time. This confusion carries over into the theoretical realm, where the chemical evolution models by \citet{hou00} predict a temporal flattening, while those by \citet{chiappini01} predict a steepening with time. 

In the present paper, we present new spectrophotometric observations of 41 PNe located toward the Galactic anticenter with galactocentric distances ranging from 0.9 to 21 kpc.  Some of these objects have no spectra available in the literature, while others have not been observed with modern detectors or over a comparably broad spectral range. These new data are used to determine the chemical abundances of He, N, O, Ne, S, Cl, and Ar in these objects in addition to electron temperatures and densities. Incorporating the information on these new objects into our previously existing abundance database of Galactic PNe \citep{hkb04, mkhs}, while adopting distances largely from a single source in the literature, provides a completely homogeneous sample of 124 PNe. We employ this database to study the Galactic chemical gradient in the disk.

Our study focuses exclusively on the determination of the oxygen abundance gradient of the Milky Way disk through the use of PNe, keeping in mind that the gradient inferred from PN studies likely applies to an uncertain mix of stars born in an earlier epoch than the gradient measured by H~II regions. While our earlier papers have looked at abundance gradients \citep{hkb04} and element-to-element ratios \citep{mkhs} for several elements in great detail, for the current study we chose instead to carry out a rigorous statistical determination of the oxygen gradient since ions of oxygen, the most abundant alpha element, are readily observed, and accurate abundance measurements can be carried out.

Our current study possesses two important characteristics which are relevant to the confidence which we will place in the outcome. First, our PN sample is completely homogeneous: we have performed all of the observations ourselves using only three instrumental systems (KPNO, CTIO, APO), all data have been carefully reduced in a uniform way, and the temperatures, densities, and abundances have been computed in a consistent fashion using the same code -- and the same atomic data and ionization correction factors -- throughout. Second, our statistical analysis accounts for uncertainties in both the oxygen abundance and the galactocentric distance of each object and uses two independent statistical packages which have been checked against each other for consistency and accuracy.

We present our new observations and the derived physical properties of the 41 new PNe, including the abundances, in Section~2. Our statistical study of the oxygen gradient is described in Section~3, while Section~4 presents our summary and conclusions. In Appendix A we compare our oxygen abundances with values from the literature.

\section{Observations and Reductions\label{obs}}
\subsection{APO Observations}
Observations of 37 PNe were carried out on the ARC 3.5m telescope at Apache
Point Observatory, NM, using the Dual Imaging Spectrograph (DIS). Each side
of the DIS has an E2V 2048 x 1024 pixel CCD. The DIS allows simultaneous
observations to be made in the blue and red portions of the spectrum,
providing the important advantages of identical slit placement and identical
sky conditions for the entire optical-near-infrared spectrum from 3600 to
9600 \AA. 

We observed with a 360" by 2" slit generally oriented E-W.  Counts on the
CCD were binned by 2 pixels (0."8) along the slit direction before readout.
The seeing was generally 1"-1."5, though for this program of extended
objects seeing has little impact except possibly for the flux calibration
star. On the blue side we used the B400 grating giving 1.83 \AA/pixel, and a
resolution of $\sim$7 \AA. On the red side we used the R300 grating,
yielding 2.31 \AA/pixel, for a resolution of $\sim$9 \AA. The spectroscopic
data are noisy between 5400 and 5700  {\AA}  where the red and blue spectra
overlap; in addition, the sensitivity of the CCD decreases at $\lambda<
3900$ {\AA} and $> 8500$ {\AA}.

Observations were obtained during five runs in 2007 between January and
November. We reduced the data in the standard fashion with
IRAF.\footnote{IRAF is distributed by the National Optical Astronomy
Observatories, which are operated by the Association of Universities for
Research
in Astronomy, Inc., under cooperative agreement with the National Science
Foundation.} Bias and overscan levels were subtracted and a normalized flat
field was applied. We checked the data for response variations in the
spatial direction along the slit and found that there was no need to apply
an illumination correction. For each PN we extracted an appropriate spatial
portion of the 2D spectrum, collapsing it to one dimension, taking
particular care to extract the same size spatial swath in the blue and the
red, since the spatial scales differ between the blue (0.40 arcsec/pixel)
and the red (0.42 arcsec/pixel) CCDs. The wavelength scale was derived from
observations of arc lamps, and the flux calibration from observations of two
or three standard stars each night. Table~1 shows our observing log.

\subsection{KPNO and CTIO Observations}

We also report here new results for four additional PNe included in the gradient analysis in \S3. These objects, observed in 2003 and 2004 with the Kitt Peak National Observatory 2.1 m telescope or the Cerro Tololo Interamerican Observatory 1.5 m telescope, are listed at the bottom of Table~1. See \S2 in Milingo et al. (2010) for details on the instrumental configurations and observing procedures. These data were also analyzed in the standard fashion using IRAF, as outlined above. 

\subsection{Line Measurements}

For each PN, the extracted spectra were combined to make a single blue
spectrum and a single red spectrum. The only exceptions arose when a strong
line was saturated in the normal-length exposures: [O~III] $\lambda$5007 in
the blue and H$\alpha$ in the red. In those cases, we obtained an additional
blue and/or red short-exposure spectrum used to measure just the line that
was saturated in the longer exposures.

We measured emission-line fluxes from the final combined, flux-calibrated
blue and red spectra of each PN. These fluxes formed the input for our
abundance determinations. We used the ELSA package (Johnson \& Levitt 2006)
for the subsequent plasma diagnostics and abundance calculations. The first
step in the analysis is to generate a table of line intensities that have
been corrected for interstellar reddening and for contamination of the
hydrogen Balmer lines by coincident recombination lines of He$^{++}$. The
reddening curves that we employed were taken from  \citet[UV]{seaton79},
\citet[visual]{savage79}, and \citet[IR]{fluks94}. ELSA iterates
calculations of the density (from [S~II] $\lambda$6717/$\lambda$6731) and
temperature (from [O~III] $\lambda$5007/$\lambda$4363) until a convergent
solution is reached, then calculates the amount of contamination to
subtract. It also calculates the correct H$\alpha$/H$\beta$ ratio for the
converged values of temperature and density. Table~2 lists the observed
fluxes and corrected intensities, including their uncertainties, along with
the calculated value of {\it c} (the logarithmic reddening parameter), the
appropriate value of H$\alpha$/H$\beta$, and the observed H$\beta$ flux
through the spectrograph slit.

\subsection{Abundances}
Along with the [O~III] temperature and [S~II] density calculated as
described above, ELSA calculates the [N~II] temperature from
$\lambda$6584/$\lambda$5755. If the required lines are observed,  the
[S~III] temperature from  $\lambda$9532/$\lambda$6312 or
$\lambda$9069/$\lambda$6312, the [S~II] temperature from
$\lambda$$\lambda$6717+6731/$\lambda$$\lambda$4068+4076, the [O~II]
temperature from $\lambda$3727/$\lambda$7323 and the [Cl~III] density from
$\lambda$5517/$\lambda$5537 are also calculated. For most of the ionic
abundance determinations to be described below, either the [O~III] or [N~II]
temperature was used along with the [S~II] density.

Ionic abundances for our sample are given in Table~3. The table shows the
emission line used to calculate each listed value of ionic abundance along
with the diagnostic temperature used in the calculation. The designation
"wm" refers to the weighted mean of the ionic abundances, weighted by the
observed flux of the line used to calculate it. Only lines flagged with an
asterisk are included in the weighted mean. At the end of each element's
list of ionic abundances in Table~3 is the ionization correction factor
(ICF). This factor represents an empirical attempt to correct the observed
ionic abundances for contributions by unobserved ionization states. The ICFs
themselves take advantage of similarities in ionization potentials among
different elements, though for some, like sulfur, there are no close
similarities, and we rely on model predictions. See, e.g., Kwitter \& Henry
(2001) for a complete discussion of the methods used to compute ionic
abundances and the ICFs used in our work.

Electron densities and temperatures for each object are presented in
Table~4, while final elemental abundances and selected element-to-element
ratios are given in Table~5. The statistical uncertainties provided in all
cases are the result of careful propagation by ELSA of line strength
measurement errors, while systematic uncertainties are not included.

In Appendix A we compare our oxygen abundances with others from the literature. We assert that although the abundance agreement is generally very good, there are some outliers.

\section{The Galactic Oxygen Gradient}

\subsection{The Determination of the Oxygen Gradient}

Table~\ref{analdata} is a compilation of information for our sample of 124 PNe that is used in the analysis to follow. For each object identified in column~1, column 2 provides the heliocentric distances taken from Cahn et al 1992 for each of the objects. All but eight of these values were available in \citet{cahn92}. References for the remaining eight objects are provided in a table footnote. The corresponding galactocentric distances in column~3 were computed using the formula given in a footnote to Table~\ref{analdata}, where we took the sun's galactocentric distance to be 8.5~kpc. Employing a distance method that was calibrated using PNe in the LMC, \citet[SSV]{stanghellini08} provide updated heliocentric distances for 101 of our 124 objects. These values and their corresponding galactocentric distances are listed in columns 4 and 5. Finally, column~6 lists the final oxygen abundances from Tables~5, using the customary form 12+log(O/H). The following analysis is based largely on the \citet{cahn92} distances, but we use the SSV results to demonstrate the effects of using different distance scales.

\begin{figure}
   \includegraphics[width=6in,angle=270]{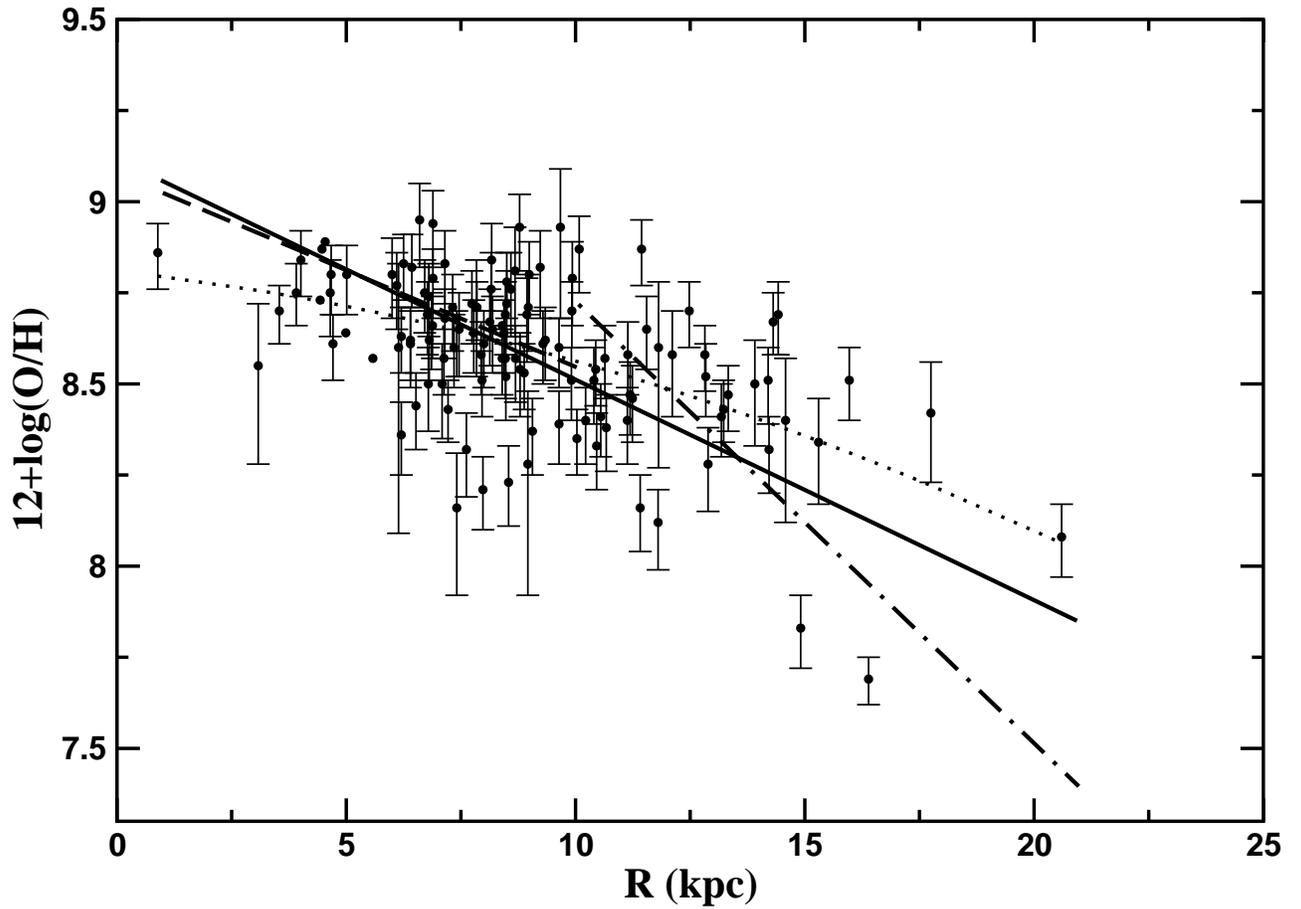} 
   \caption{12+log(O/H) versus galactocentric distance, R, in kpc for PNe in our total sample. Least squares fits are indicated by line type for the total sample (solid), objects for R$<$10kpc (dashed), R$>$10kpc (dot-dashed), while a quadratic fit is shown with a dotted line.}
\label{ovr_pne}
\end{figure}

Fig.~\ref{ovr_pne} is a plot of 12+log(O/H) versus galactocentric distance (column~3, Table~\ref{analdata}) in kpc for the 124 objects in our PN sample with well-measured distances. The PN sample is a compilation of objects from \citet{hkb04}, \citet{mkhs}, plus those objects reported upon for the first time in this paper (HK10). Our sample contains only those objects classified as either Type I or Type II, i.e., objects whose location and kinematics indicate that their progenitors were members of the disk population. We use the heliocentric distances from \citet{cahn92} from which we calculate galactocentric distances, except for the following objects (heliocentric distance sources in parentheses): H2-18, He2-48, He2-55 \citep{maciel84}; IC 418 \citep{guzman09}; M3-15 \citep{zhang95}; NGC 6720, NGC 6853, NGC 7293 \citep{harris06}; H3-75 \citep{amnuel84}; K3-64, K3-93 \citep{jaskot08}; St3-1 \citep{tajitsu98}.

Our primary goals are: (1) to subject the data to careful statistical tests in order to determine the nature of the oxygen abundance gradient in the Galactic disk as measured by PNe; and (2) to compare our result with analogous studies using other types of probes. In order to be included in the gradient analysis, we required that a PN have good determinations of distance, plus good  $T_e$[O~III], $T_e$[N~II], and $N_e$[S~II], which means reliable flux measurements of $\lambda$4363, $\lambda$5755, $\lambda$$\lambda$6717/6731. Combining the 41 PNe whose measurements are presented here with additional disk PNe already in our database, 124 objects qualified for inclusion in the sample on which we performed a thorough linear regression analysis. 

Our first step was to gauge the strength of a correlation between O and $R_g$ by computing a Pearson correlation coefficient, r, for the entire sample of objects, using the program {\it pearsn} in \citet{press03}. The Pearson correlation coefficient is the ratio of the covariance of two variables--in this case O and R$_g$~--~to the product of their individual standard deviations. The square of this coefficient, r$^2$, called the coefficient of determination, is a measure of the strength of the linear relationship between the two variables. Both r and r$^2$ are invariant under linear transformation, and thus uncertainties in O and $R_g$ are irrelevant. [See \citet{rodgers88} and \citet{hays81} for a more detailed description of the correlation coefficient.] 

For our complete PN sample we found that r=-0.54 and thus r$^2$=0.29. The proper interpretation of this result is that 29\% of the variability in 12+log(O/H) (henceforth O) is accounted for by R$_g$ under the assumption that a linear model exists which describes the O-$R_g$ relation. The same statistical program also predicts the probability that this result could arise from a completely uncorrelated parent population. Thus, if the null hypothesis is that O and R$_g$ are uncorrelated in the parent population, then the probability that a sample of 124 objects drawn from such a parent population will have a value of $|r|  \ge$0.54 is only $3.3 \times 10^{-11}$. In other words, there is a vanishingly small probability that this null hypothesis is being falsely rejected (Type~I error; Hays 1981), and so the correlation appears to be truly nonzero.

Our next step was to derive a good linear model, i.e., least squares fit, for the data in the O-$R_g$ plane. We did this by using the program {\it fitexy} in \citet{press03}, which accounts for errors in both coordinates: O and R$_g$. Our initial attempt included the one-sigma uncertainties in O taken directly from our published estimates, while for R$_g$ we {\it assumed} a standard one-sigma uncertainty of $\pm$20\% \citep{stanghellini06}. These factors resulted in a slope, b, of -0.066$\pm$.0055 dex/kpc and an intercept, a, of 9.15$\pm$.04. Assuming that the errors are normally (Gaussian) distributed, this fit has an associated $\chi^2$ value of 178.0 with 122 degrees of freedom $\nu$ (124 sample objects minus 2 determined parameters, a and b), or a reduced $\chi^2$ ($\chi^2_{\nu}=\chi^2/\nu$) of 1.46. 

Ideally, we would like the value of $\chi^2_{\nu}$ to be unity; values greater than this suggest that the random errors have been underestimated (or there is an unidentified factor that results in the scatter being greater than what is described by the uncertainties). Meanwhile, values less than unity indicate that one has overestimated the errors. In the spirit of Goldilocks, when $\chi^2_{\nu} \approx 1$, then the assumed uncertainties satisfactorily explain the scatter of the data points around the model. A method of quantifying this idea is to compute the probability that a value of $\chi^2_{\nu}$ greater than or equal to the one determined from the data could result from randomly sampling the parent population described by the model and for the same number of degrees of freedom \citep{bevington03}. This is done by integrating the $\chi^2$ probability function from $\chi^2_{\nu}$ to infinity to obtain $q_{\chi^2}$, often referred to as a goodness-of-fit parameter. A relatively large $q_{\chi^2}$ suggests that within the confines of the one-sigma errors, $\chi^2_{\nu}$ determined from the data  has a reasonable chance of occurring and the model is likely adequate in characterizing the data. Likewise, a relatively small value means that the errors cannot explain $\chi^2_{\nu}$ and that the model is likely inadequate. The threshold value for $q_{\chi^2}$ in statistical studies is usually taken to be 0.05 \citep{hays81}. In our current case, q$_{\chi^2}$ was determined by the program {\it fitexy} to be 0.00074, and so we deemed this particular trial model unsatisfactory.

Before continuing further with our analysis, we decided to write our own program for computing linear least-square fits in order to verify the accuracy of the routines in \citet{press03} as well as to understand the methods better. Therefore, we wrote a program in {\it Mathematica} to generate a least-squares fit of data with errors in both coordinates. Specifically, we solved iteratively the exact ``least-squares cubic'' derived by \cite{York66,York69} as unified by \cite{YEM04} \citep[see also][]{MT92}. To verify our procedures and implementation, and assess the accuracy of the resulting fitting parameters we successfully checked two test cases: (1) the classic data case developed by \citet{Pearson01} with errors assigned to both coordinates by \cite[Table II]{York66} and previously considered by~\cite{Lybanon84,Lybanon85,PM72,Reed92} and~\cite{YEM04}; and (2) data for electron temperatures of ionized hydrogen in the Magellanic clouds derived by~\cite{VvdH02} from abundances of O$^{+2}$ and S$^{+2}$ and previously considered by~\cite{YEM04}.

In attempting to produce a fit to our own data sample of 124 objects, we assumed the same one-sigma errors as in the above trial with {\it fitexy}. Only six iterations were required to determine the slope and intercept to 10~decimal places; the rounded values are $b=-0.066\pm.006$ and $a=9.15\pm.04$, a $\chi^2$ value of 178.0 and a $\chi^2_{\nu}$ of 1.46. The calculated standard deviations in the slope and intercept (the ``standard errors'') were computed following the guidelines in \citet[Chap. 9]{Cowan98}. Since our results are in excellent agreement with those produced by the routines in \citet{press03}, we confidently continued our analysis using the latter formulations. 

We sought to improve the fit by scaling the uncertainties of both O and $R_g$ upward independently, since increasing the standard deviation of either or both coordinates reduces the size of $\chi^2$; correspondingly, the value of $q_{\chi^2}$ goes up. For each PN the new uncertainty $\sigma_O = \sigma_O' \times f(O)$, where $\sigma_O'$ is the originally-determined uncertainty in O and f(O) is the scaling factor. For $R_g$, we set $\sigma_{R_g} = R_g  \times f(R_g)$. Thus, we experimented with different values of the scaling factors f(O) and f($R_g$) in systematic fashion in order to find reasonable ranges in these numbers for satisfactory fits. Values of f(O) ranged between 1 and 2, while values of f($R_g$) ranged between 0.0 and 0.3. 

The test results for each combination of scaling factors are given in Table~\ref{lsf}, where the first two columns list the scaling factors, the third column lists the resulting slope of the fit, and the fourth column provides the value for the q$_{\chi^2}$, with the rows ordered top to bottom by increasing q$_{\chi^2}$. We see that acceptable models correspond to gradient slopes ranging between -0.041 and -0.074~dex/kpc. Additional tests not reported here indicated that larger values of  f($R_g$) will push the lower limit downward but it is not likely to go below -0.08. However, realistic values of f($R_g$) likely do not greatly exceed 0.3, based upon the brief discussion in \citet{stanghellini06}. 

To refine our model further, we assumed that $ f(R_g) =0.2$, consistent with the value recommended by \citet{stanghellini06}, and then adjusted the value of f(O) until the fit explained all of the scatter in O, i.e., $\chi^2_{\nu}$=1.0. This criterion was met when f(O)=1.40, at which point the resulting model had the following parameters: a=9.09$\pm$.05 dex, b=-0.058$\pm$.006 dex/pc$^2$, $\chi^2$=121.8, $\chi^2_{\nu}$=1.00, and $q_{\chi^2}=0.49$. Since the last statistic is well above the threshold of 0.05, the model cannot be rejected, and we judge the fit to be acceptable. These results for the total PN sample are shown in the first row of Table~\ref{gradients}, where the first column describes the sample or subsample under consideration, the second column provides the number of objects in the sample, and the next seven columns give the y intercept, slope, $\chi^2$, reduced $\chi^2_{\nu}$, goodness-of-fit parameter, correlation coefficient, and the correlation probability factor, respectively. (Subsequent rows provide the statistics for PN subsamples or other object types.) 

Thus, we adopt the following analytical behavior of the oxygen abundance with galactocentric distance (Table~\ref{analdata}, columns 3 and 6) as that which best describes the data for the total sample presented in Fig.~\ref{ovr_pne}:
\begin{equation}
12+log(O/H) = (9.09\pm.05) -(0.058\pm.006) \times R_g\label{e1}.
\end{equation}
This regression model is indicated with a solid bold line. At the Sun's distance from the Galactic center, 8.5~kpc, our model predicts a value of 8.60$\pm$.07 for 12+log(O/H), close to a recently-determined solar value of 8.69$\pm$.05 by \citet{asplund09}.

We checked to make sure that our inferred gradient in eq.~\ref{e1} was not being influenced too heavily by the few extreme points at large $R_g$. To do this, we started with the object with the greatest galactocentric distance and excluded PNe one by one, calculating a new value for the gradient each time. All resulting gradients obtained in this manner were slightly flatter -- the maximum slope was -0.051, occurring after eliminating all of the outer six objects. Thus, we feel confident that the model in eq.~\ref{e1} is not being influenced unduly by the few objects in our sample located at large radial distances.
 
The need to increase the value of $\sigma_O'$ by 40\% beyond our previously established level of uncertainty in order to reach $\chi^2_{\nu}=1$ suggests that objects in our PN sample possess some natural scatter which is related to real abundance differences among them. Such differences could be caused by one or more of the following: (1)~age differences of the progenitor stars, where the older ones formed out of less metal-rich material than younger ones; (2)~inhomogeneous mixing of the interstellar material out of which the progenitors formed; (3)~stellar diffusion, in which stars migrate along the disk from a galactocentric radius where they formed to their present location; and/or (4)~systematic errors in the abundance determination process. It is clear from Table~\ref{lsf} that the value of the gradient is sensitive to uncertainties in $R_g$ as well as to the oxygen abundance. However, our adopted standard uncertainty of 0.20 for $R_g$ is considered to be quite reasonable (L. Stanghellini, private correspondence), and thus we still maintain that at least some natural scatter in the oxygen abundances exists.  In any case, it is beyond the scope of this paper to further discuss the origin of the natural scatter.

For comparison purposes and to see the effects of using a different set of distances, we used the 101 objects in Table~\ref{analdata} for which SSV provided heliocentric measurements in order to repeat the regression statistics computation. We first compared directly the two distance sets and found that while most objects fell very close to a line of 1:1 correspondence, roughly 10\% of the objects did not. Interestingly, these differences turned out to have a noticeable impact on the regression results when performed with the SSV distances. Using the same uncertainties that produced the result in equation~\ref{e1}, i.e., f(O)=1.4 and f(R$_g$)=0.2, the SSV distances now imply a slope of -0.042$\pm$.004 with $\chi^2_{\nu}$=1.40. Additional tests involving larger (probably unrealistic) values of  f(R$_g$) to drive $\chi^2_{\nu}$ down resulted in a further flattening of the slope. We shall comment on this result at the end of this subsection.

Next, we subdivided our sample by galactocentric distance, Peimbert type, and height above the Galactic plane. In the first two cases we computed a least squares fit and correlation coefficient for each subgroup to look for distribution differences among them.
 
We divided objects into two groups based upon whether their galactocentric distance placed them outside or inside the 10~kpc circle. This distance was chosen because it coincides with the position of the sharp discontinuity in [Fe/H] reported by \citet{twarog97} in which the mean value of [Fe/H] beyond this point was found to drop by 0.3~dex. We then computed the regression statistics separately for the two groups. PNe out to 10~kpc have a gradient and correlation coefficient of -0.054$\pm$.013 and -0.21, while those beyond this distance have values of -0.12$\pm$.14 and -0.39, respectively, and their curves are shown in Fig.~\ref{ovr_pne}. In the same figure we also show a quadratic fit to the data which also indicates a steepening of the gradient, where the function corresponding to this curve is $12+log(O/H)=8.81-0.014 \times R_g - 0.0011 \times R_g^2$. The quadratic fit predicts a relatively flat gradient at small galactocentric distances but a rather steep one past 10~kpc. Although these results suggest that the gradient steepens in the outer disk, we caution that the scatter is broad beyond 10~kpc, and only additional data at these distances will allow us to reliably determine the true behavior of the slope. Interestingly, some authors present empirical evidence for a {\it flattened} gradient in the outer disk \citep{vilchez96,costa04,maciel08,pedicelli09}. For example, \citet{costa04} use PNe to estimate that the oxygen gradient is -0.09~dex/kpc between 4 and 5~kpc, but the slope becomes flat at 11~kpc and beyond. Likewise, \citet{pedicelli09}, using a sample of 265 Cepheids between 5-17 kpc from the Galactic center, find the iron gradient to be three times steeper inside the 8~kpc circle than outside of it. This disagreement is also apparent in published model results: All of the chemical evolution models by \citet{fu09} and the models employing a constant star formation efficiency by \citet{marcon10} predict a steepening gradient in the outer disk, while the models by \citet{marcon10} which assume that the star formation efficiency falls off with radial distance, predict a flattening of the gradient.

\begin{figure}
   \includegraphics[width=6in,angle=270]{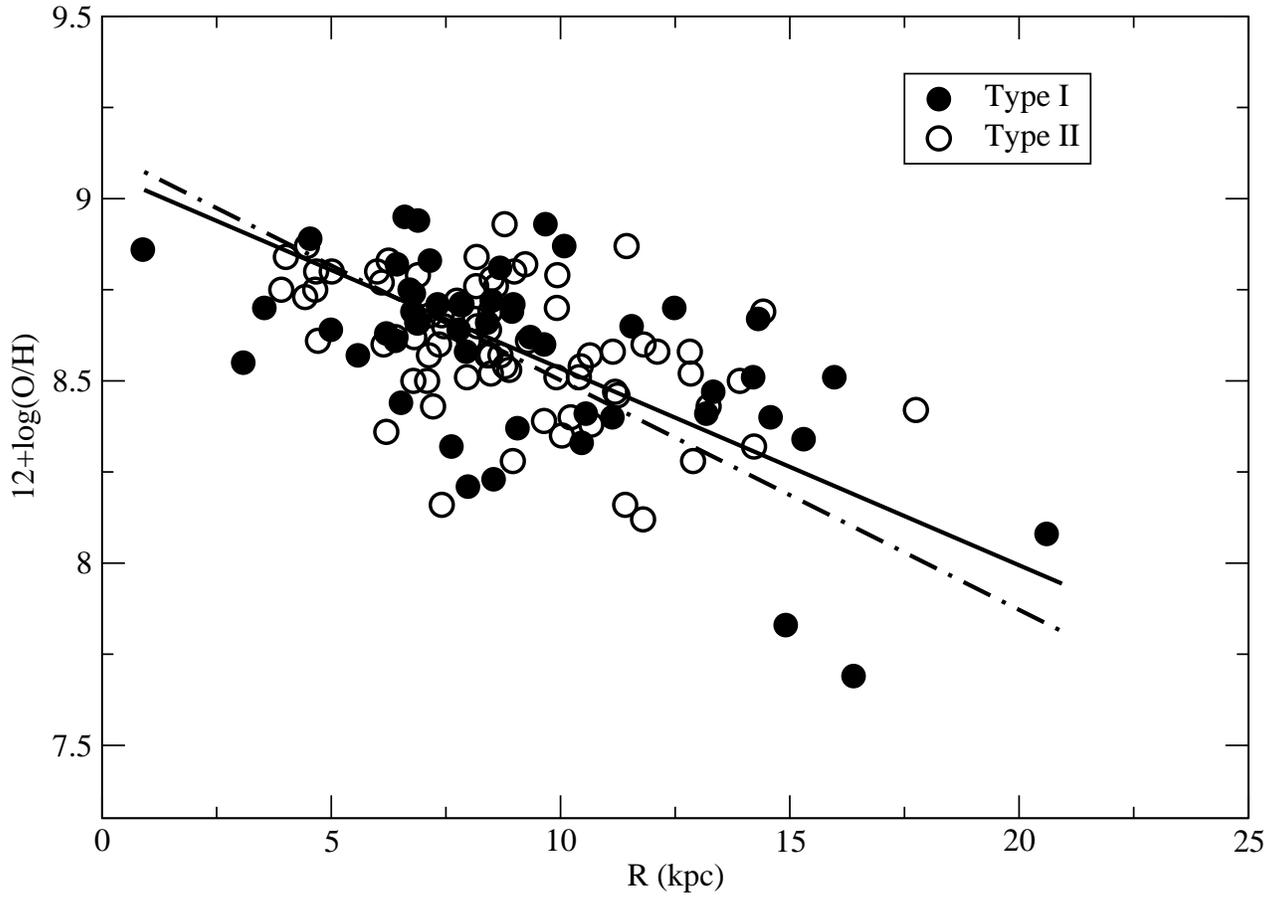} 
   \caption{12+log(O/H) versus galactocentric distance, R, in kpc for objects in our sample. Type I PNe are symbolically distinguished from Type IIs as shown in the legend. Least squares fits for Type I and Type II are shown with dot-dashed and solid lines, respectively.}
\label{ovr_type}
\end{figure}

Figure~\ref{ovr_type}  shows the distributions by Peimbert type along with the regression lines, while the complete statistical information is provided in Table~\ref{gradients}. The gradient slopes (and uncertainties) were found to be -0.061$\pm$.008 and -0.053$\pm$.010 for Type I and Type II, respectively, while the correlation coefficients were -0.59 and -0.44, respectively. Thus, within the uncertainties the two types appear to be indistinguishable, a conclusion supported visually in Fig.~\ref{ovr_type}. In addition, the intercept values are also comparable, and so we are unable to see any statistically significant difference in oxygen abundance patterns between the two PN types. This conflicts with the claim of a gradient difference between PN types by  S10, who find a slope of -0.035$\pm$.024~dex/kpc for Type~I PNe and a slope of -0.023$\pm$.005~dex/kpc for Type~II PNe. These authors continue by relating PN type with progenitor age and conclude that the gradient has steepened with time. Interestingly, their conclusion conflicts with that of \citet{maciel08}, who claim that the gradient has flattened over time. Resolving the problem of temporal slope change is extremely important from the standpoint of chemical evolution of the MWG, but more than likely the situation is currently being complicated by small-number statistics along with real scatter in the oxygen abundances of objects at similar galactocentric distances.

\begin{figure}
   \includegraphics[width=6in,angle=270]{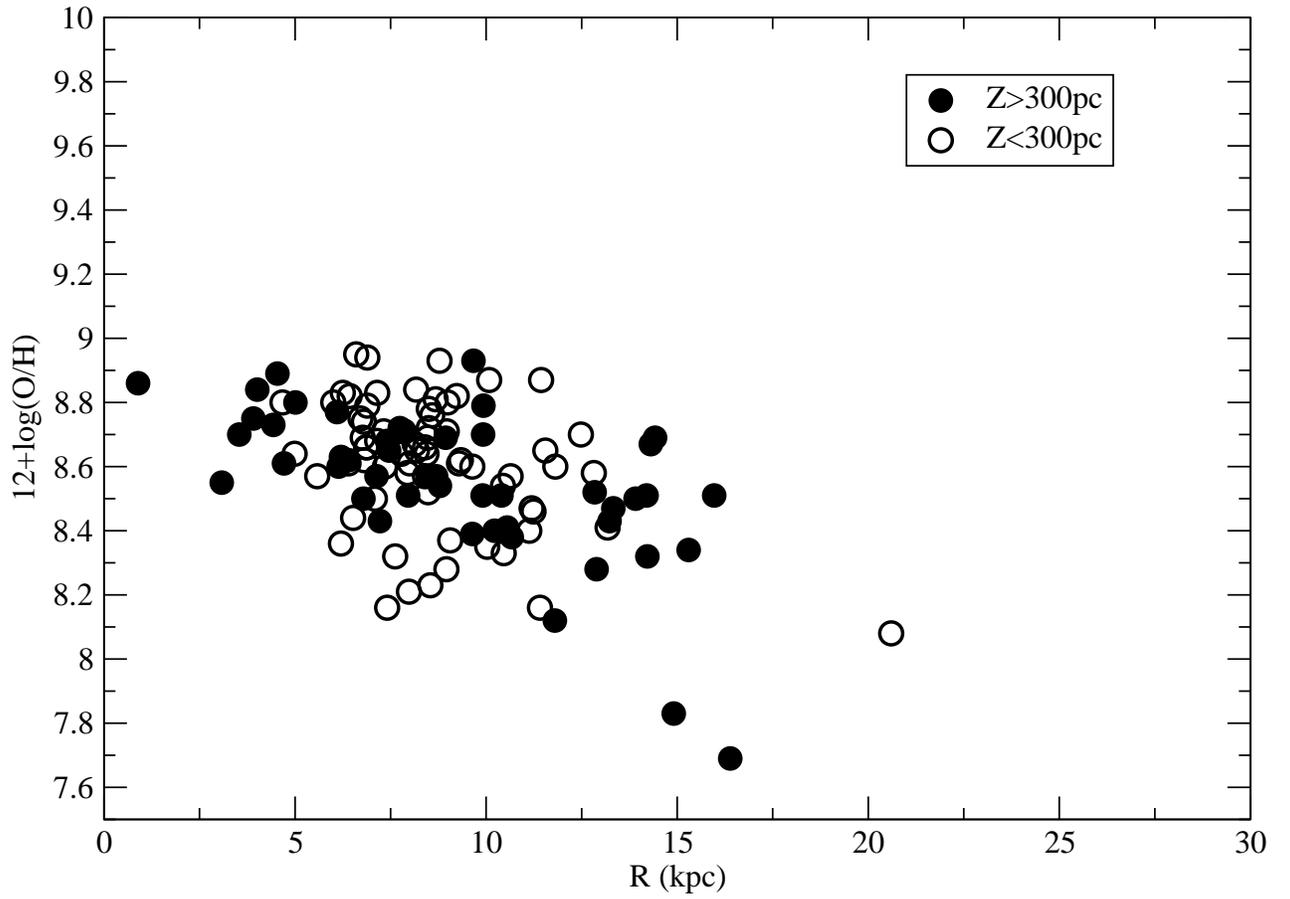} 
   \caption{12+log(O/H) versus galactocentric distance, R, in kpc. PNe are separated according to whether they are greater than or less than 300 kpc from the Galactic plane in the vertical direction.}
\label{ovr_z300}
\end{figure}

Figure~\ref{ovr_z300} separates the sample by vertical distance from the Galactic plane, where we adopt Z=300 pc, the general disk population scale height \citep{cox00}. The gradient slopes were found to be -0.038$\pm$.007 and -0.033$\pm$.009 for objects above and below this height, respectively, while the correlation coefficients are -0.63 and -0.39, respectively. These numbers, along with a visual inspection of the lines in the figure, indicate that the distributions of the two subsamples are indistinguishable.

We repeated the above exercise but this time adopted Z=100 pc, the Population I scale height \citep{cox00}. The regression analysis inferred slopes of -0.038$\pm$.006 and -0.013$\pm$.028 for PNe above and below 100 pc respectively, while the correlation coefficients are -0.59 and -0.09, respectively. Here the differences, particularly in the correlation coefficients, are likely the result of the relatively small galactocentric distance range for those objects within 100 pc of the plane.

Finally, eq.~\ref{e1} was used to normalize all oxygen abundances to the same galactocentric distance and then these adjusted oxygen abundances were plotted against the corresponding z~distance perpendicular to the Galactic plane. While we do not show the plot here, we can report that no correlation was found. We conclude that within our database there is no evidence for a correlation between oxygen abundance and vertical distance from the plane.

\begin{figure}
   \includegraphics[width=6in,angle=270]{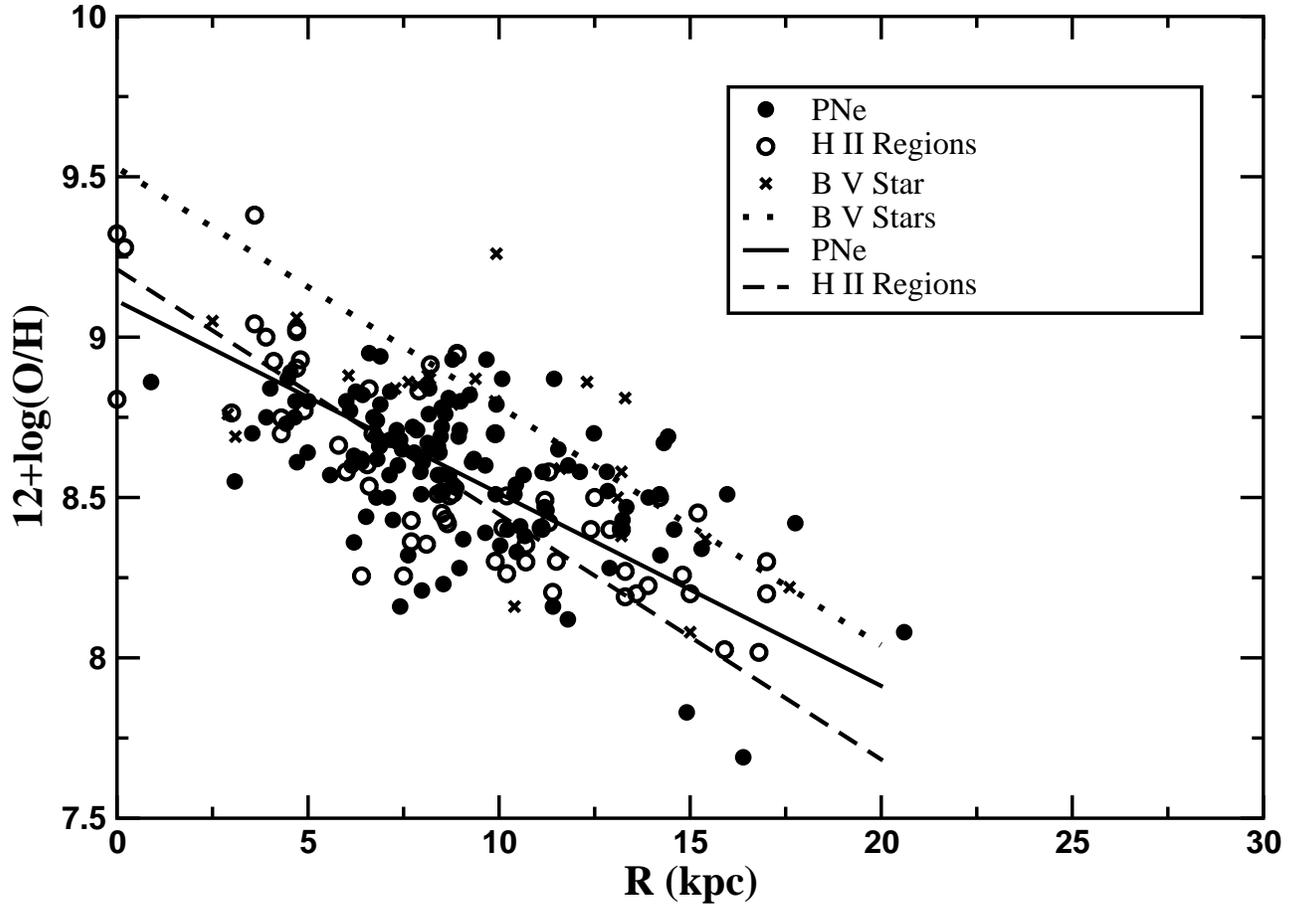} 
   \caption{12+log(O/H) versus galactocentric distance, R, in kpc. Planetary nebulae are indicated with filled circles, while H II regions are indicated with open circles, and stars are indicated with stars. The lines are least squares fits to the respective data types, as indicated in the legend. The sources for the samples of PNe, H~II regions, and B~V stars are the same as those provided in Table~\ref{gradients}.}
\label{ovr}
\end{figure}

Table~\ref{gradients} also provides comparisons of our derived PN oxygen abundance gradient for the disk of the MWG with analogous results for what are likely to be younger populations of H~II regions and B V stars. The bottom two rows of Table~\ref{gradients} show the results of applying our statistical methods to an H~II region and a B V star sample which we compiled using abundances taken directly from the literature. The H II region sample was compiled from measurements reported in \citet{afflerbach97}, \citet{vilchez96}, and \citet{deharveng00}, and covers the galactocentric distance range of 3 to 17 kpc. We used the abundance uncertainties as quoted in the papers and adopted a standard error in galactocentric distance of 2~kpc for an acceptable fit. Likewise, the B~V star sample was compiled from studies by \citet{smartt01} and \citet{rolleston00} and extends from 2-18 kpc. In this case it was necessary to scale the abundance uncertainties by 1.9 in order to achieve a suitable fit, while the galactocentric distance errors were taken directly from the papers. We present a comparison of the total PN, H~II region, and B~V star samples listed in Table~\ref{gradients}, along with their corresponding least squares fits, in Figure~\ref{ovr}.

 In both the H~II region and stellar studies we see strong inverse correlations between oxygen abundance and galactocentric distance and good agreement in slope values with that of our total PN sample. Note that the correlation coefficients in both cases indicate the presence of a stronger correlation than for our total PN sample, although the gradients are consistent. Interestingly, this is despite the fact that there is likely to be a mean age difference between our PNe and the H~II region and B~V samples, as the latter two generally represent a younger population than the first group. 

Table~\ref{gradientscom} presents a broad comparison of our adopted gradient (eq.~\ref{e1}) for our total sample with many other gradients from PN studies along with H~II region, Cepheid, and  B~V star studies. In contrast to the gradients listed in Table~\ref{gradients}, which we computed directly from the published data, the gradients in Table~\ref{gradientscom} are those provided by the authors. For the source identified in column~1, the second column gives the number of objects in the sample, while the next three columns provide the gradient in dex/kpc, the  type of objects in the sample, and the galactocentric distance range covered, in kpc. 

First, we see a wide range of gradient values among the PN samples listed. In particular, the difference between this paper and S10 is interesting, since both samples contain a large number of objects and cover a broad range in galactocentric distance. Clearly, the gradients inferred by these two studies are statistically different, with the S10 paper proposing a flatter gradient than the one we find here. We shall expand on the meaning of this difference in the next subsection. In the meantime, we verified that the difference in the linear models derived by S10 and us were not attributable to the use of different oxygen ICFs. We recomputed our abundances using their ICF formula for oxygen, taken from \citet{kb94}, and found essentially no difference in either slope or intercept.

The other three PN-derived gradients by \citet[H04]{hkb04}, \citet[C04]{costa04}, and \citet[P06]{perinotto06} cover slightly more restricted distance ranges. H04 used their own spectrophotometric observations to derive abundances of Types I and II PNe in the MWG disk; C04 likewise used their own spectral measurements to derive abundances of 26 Galactic anticenter disk objects and combined these with PNe from the sample of \citet{maciel99} to infer their gradient; and P06 compiled observations from the literature of Type~II PNe but reprocessed all line strengths themselves to produce a homogeneous abundance set for determining their gradient. The gradients of S10, H04, and P06 appear to be very similar, given the uncertainties associated with each.

The remaining gradients in Table~\ref{gradientscom} include those derived from optical and IR studies of H~II regions by \citet[R06]{rudolph06}, Cepheids by \citet[P09]{pedicelli09}, and B~V stars by \citet[SR97]{smartt97}. (Note that P09 report an Fe/H gradient.) Gradients in these four cases are consistent with those for PNe, although the uncertainties are somewhat larger in the former.

In summary, we see strong evidence once again for a negative oxygen gradient in the disk of the MWG. However, there is no evidence in our own work that the slope differs between PN Types I and II or between objects of different vertical distance from the Galactic plane. Because of the broad abundance spread beyond 10~kpc, we cannot confirm with any confidence the claim by many others that the gradient changes (either steepens or flattens) in the outer regions of the disk. {\it We therefore adopt the simplest hypothesis, to wit, for now that the gradient is constant along the disk's entire length.} 

The most troubling aspect of our gradient derivation exercise is that both the uncertainties in the distance scale and the oxygen abundances prevent us from pinning down the value of the slope at this time to an accuracy better than about 0.02~dex/kpc. This became apparent above when we substituted the SSV distances for the ones used in the original analysis, i.e., column~2 of Table~\ref{analdata}, and were forced to add more natural scatter to the oxygen abundances in order to reduce the value of $\chi^2_{\nu}$ to an acceptable level. From our complete exercise we consider it very likely that the true slope is within the range of -0.04 to -0.06 dex/kpc, but we cannot refine the number beyond that point. Essentially we have reached a confusion limit regarding the abundance gradient as derived using PNe. Perhaps with the launch of the GAIA probe\footnote{The main goal of the Gaia mission, as stated at the GAIA website, ``is to make the largest, most precise three-dimensional map of our Galaxy by surveying an unprecedented one per cent of its population of 100 billion stars.''} in 2012 trigonometric distances to many Galactic PNe can be measured with much better accuracy, and the uncertainty introduced by the distance scale for PNe can be significantly reduced.

\subsection{Gradient Uncertainty and Theoretical Models}

Given the range in values for the slope of the Galactic disk oxygen gradient displayed in Table~\ref{gradientscom}, the obvious uncertainties related to an unsettled distance scale, and the extensive work that is currently going on to reconcile these differences, it is appropriate to ask whether continuing our efforts to better define the characteristics of the oxygen abundance distribution is likely to offer much in the way of additional improvement in our understanding of the chemical evolution of the MWG disk. To address this question, we consider the model-predicted sensitivity of the Galactic oxygen gradient to a few parameters related to galactic chemical evolution. 

Recent detailed chemical evolution models of the Milky Way disk by \citet{marcon10} illustrate how the value of the present day gradient as reflected by studies of H~II regions is influenced by the threshold density for star formation, i.e., the surface density above which star formation occurs but below which it doesn't, and the variation of the star formation efficiency, i.e., the ratio of the mass of stars formed to the mass of gas available for forming them. Their models predict that between 4-14~kpc in galactocentric distance, when the star formation efficiency is held constant and the star formation threshold is reduced from 7 to 4 M$_{\odot} pc^{-2}$, the predicted oxygen gradient flattens, going from -0.059 to -0.025 dex/kpc. Note that these gradient values are almost exactly the same as those reported in this paper (-0.058) and the one by S10 (-0.023), respectively, for our total samples. At the same time, holding the threshold constant at 4~M$_{\odot}pc^{-2}$ but allowing the star formation efficiency to be a function of galactocentric distance produces little effect. Similar patterns are seen over more restricted ranges in R$_g$.  Therefore, these models indicate that improving our knowledge of the oxygen gradient in the Galactic disk will allow us to pin down the value of at least for one important parameter, the star formation threshold.

In a similar fashion, detailed chemical evolution models of the Milky Way disk by \citet{fu09} tested the effects of the infall timescale, the star formation law, and disk formation timescale (DFT) on the size of the abundance gradient. In their models the disk is assumed to begin forming out of matter infalling from the halo at a time specified by the DFT, while the rate of infall decreases by $e^{-1}$ over a time defined by the infall timescale. At the same time, the star formation rate is proportional to some power of the local gas surface density and may also be a function of radial distance along the disk.

\citet{fu09} tested the impact of these three parameters on the steepness of the present day abundance gradient and found that the DFT had the largest effect, while the other two parameters influenced the gradient in a minor way only. For example, when DFT is zero (disk formation begins at the moment of Galaxy formation) the gradient ranged in value from -0.009 to -0.027 dex/kpc. However, when the DFT value increases radially the resulting gradient range steepens to -0.056 to -0.091. The range in values in each case is the result of changing the star formation law or the infall timescale. The point here, of course, is that as in the case of the \citet{marcon10} models we see  a predicted range in gradient steepness that corresponds roughly to the observed range in the oxygen gradient presented in Table~\ref{gradientscom}.  

Thus, chemical evolution models strongly indicate that some important parameters of chemical evolution of the disk, the density threshold for star formation and the timescale for disk formation, are expected to have a measurable influence on the value of the abundance gradient. Therefore, continuing to refine our quantitative description of the gradient is clearly worthwhile. Further studies should continue to focus on extending the baseline out to a radial distance beyond 20~kpc from the Galactic center in order to firmly establish the character of the gradient in terms of its actual value in addition to its shape, since there is observational support for a gradient of constant slope plus evidence for gradients which steepen or flatten with increasing galactocentric distance.

\section{Summary/Conclusions}

We have obtained new spectrophotometric measurements of 41 planetary nebulae located in the direction of the MWG anticenter using the facilities of the Apache Point Observatory. Density, temperature, and abundance information for each object were subsequently derived using the software package ELSA. These results were in turn added to our large database of homogeneously determined abundances in order to study the oxygen abundance gradient in the Galactic disk. The complete PN sample comprised both Type I and II objects, 124 nebulae in all, extending in galactocentric distance from 0.9 to 21~kpc, where all observations, data reductions, line measurements, and abundance computations were carried out using exactly the same methods and executed by the same personnel throughout.

The primary focus of this paper (in addition to presenting new spectroscopic data and abundances for 41 anticenter PNe) was to carefully study the nature of the oxygen abundance distribution across the Galactic disk. To accomplish this, we computed linear regression models (least squares fits) of the total sample along with subsets by including uncertainties both in the abundances themselves and in the galactocentric distances of the objects. We have concluded the following:

\begin{enumerate}

\item Our total sample of PNe is well fit by a linear regression model whose form is $12+log(O/H)=(9.09\pm.05) - (0.058\pm.006) \times R_g$ with a correlation coefficient of -0.54, where R$_g$ is galactocentric distance in kpc. In computing this model we accounted for uncertainties in both the dependent and independent variables. Related probability calculations for testing the strength of these parameters support the validity of our results. A parallel analysis using the recent distances for a smaller sample from SSV resulted in a oxygen abundance slope of -0.042$\pm$.004, i.e., somewhat flatter than the above results. 

\item We find clear evidence for natural scatter in oxygen abundance for our sample PN population, scatter well beyond what can be explained by observational (statistical) uncertainties.

\item Because of the problems related to distance scale uncertainties and real scatter in oxygen abundances, we propose that gradient determination studies involving PNe have reached a confusion limit, and that further resolution will only come with improved PN distances and understanding of the natural scatter in oxygen abundances which appears to exist.

\item We see no obvious indications that Type~I and Type~II PN oxygen abundances are characterized by different gradient slopes. Likewise, there seems to be no correlation between an object's oxygen abundance and its vertical distance from the Galactic plane.

\item We find some evidence that the oxygen gradient beyond 10~kpc is steeper than it is inside this distance. In addition, both the first and second order parameters of a quadratic fit to our entire sample are negative, again suggesting that the gradient steepens with increased galactocentric distance. However, we feel that given what appears to be natural scatter in the abundance data along with the presence of only a relatively small number of sample objects beyond about 17~kpc in the database, it is too early to conclude that the gradient is anything but linear with a single slope across the entire disk.

\end{enumerate}

Concerning the second point above, in order to achieve the fit described in point~1, we were forced to assume one-sigma abundance uncertainties that were 40\% greater than the propagated ones; without this alteration, the $\chi^2$ value was too large. This result suggests the presence of real scatter in oxygen abundance among PNe at the same radial distance from the Galactic center. We suggested a few potential theories to explain this finding.

Finally, we briefly considered the importance of achieving a consensus on the gradient value of the oxygen abundance distribution across the Galactic disk, as the current range in published slopes insufficiently constrains chemical evolution models and thus our understanding of how the gradient originally formed and subsequently evolved. Related to this point is the need to clarify the actual {\it form} of the fitted model, e.g., linear, quadratic, discontinuous with multiple slopes, etc. The problem in doing this right now appears to be twofold. First, there is a paucity of oxygen measurements beyond 15~kpc from the Galactic center. Clearly, extending our sampling out farther will help to develop a consensus regarding the form of the abundance behavior, both qualitatively and quantitatively. But second, there is the problem of inherent, real scatter in the oxygen abundances whose cause is unknown but whose presence vastly complicates the study of the disk's oxygen distribution. This situation makes the study of the abundance gradient more difficult, of course, but it also brings to light a problem whose solution is bound to tell us something very interesting about how the level of interstellar oxygen in the Galaxy evolves.

\acknowledgments

We wish to thank the staffs of the Apache Point Observatory and of KPNO and CTIO for their generous assistance during the observational phase of this project. We thank Steve Souza for his assistance at the
telescope during the observations at KPNO and CTIO. We also appreciate the help of Joe Rodgers and Eric Abraham regarding the methods for the statistical analysis of our data. RBCH, KBK, BB, and AJ also thank NSF for their generous support through grants AST-0806577 to the University of Oklahoma, AST-080490 to Williams College, and AST-0880201 to the University of Washington. Finally, we are grateful to the referee for suggesting numerous changes which improved the paper.

\appendix

\section{Abundance Comparison}

Our published work (including both the new objects reported here and others we have reported on elsewhere) includes 10 PNe in common with \citet{wlb05}. Our oxygen abundances for these 10 PNe agree with theirs on average to within 2\%, at the worst differing by 17\%. \citet{costa04} report on 11 objects in common with our total sample. Though their values are almost always higher than ours, agreement is reasonable for most of the objects, but for three (A12, M1-13, and YC2-5) the difference exceeds a factor of two. Perhaps this is due to their having to rely on [O~II] $ \lambda$7323 for the O$^+$ abundance, and not being able to measure $\lambda$3727: for nine of these 11 objects (excluding for M1-6, where we agree within 25\%, and YC2-5, a high-ionization PNe with little O$^+$) the observed flux of $\lambda$3727 is much stronger than  that of $\lambda$7323.

\begin{figure}
   \includegraphics[width=6in,angle=270]{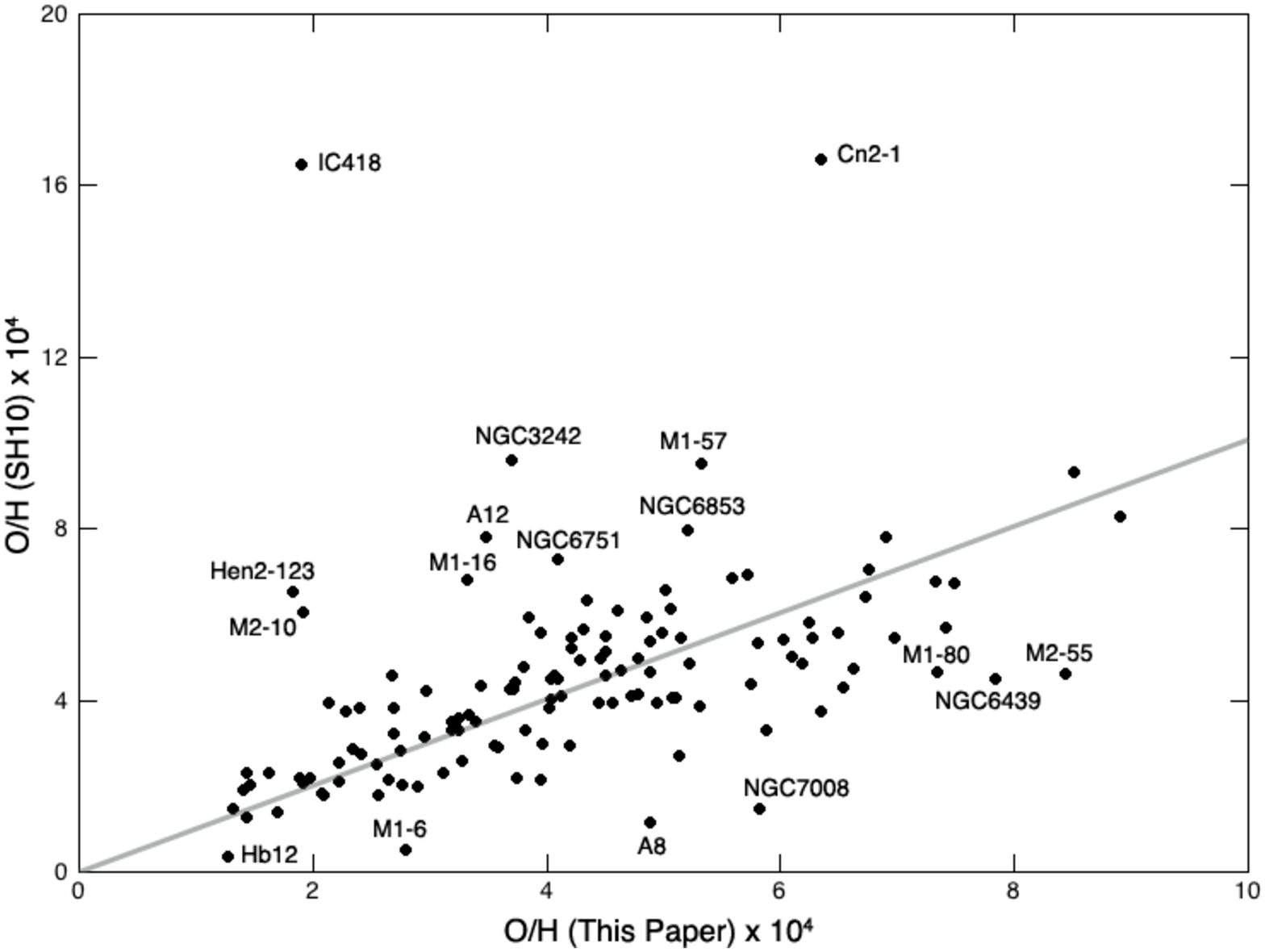} 
   \caption{O/H x 10$^4$ abundances from \citet[vertical axis]{stanghellini10} and the present sample (horizontal axis) for objects common to both. The 1:1 track is shown with a straight line.}
\label{ocomparisons}
\end{figure}

Fig.~\ref{ocomparisons} compares the oxygen abundances (O/H) for our entire published sample with those in S10 for the 126 objects in common. The gray line is the unity line. In general, the values lie within $\pm$50\% of the unity line; outliers are identified by name. In general, the points at low O/H scatter around the unity line; at higher O/H our values tend to be larger than SH10's. Since SH10 do not explicitly list the abundance reference for each nebula, we cannot, in fact, trace the sources of the observed discrepancies. However, we can speculate on one possible cause, especially for highly-ionized nebulae: it could be the ionization-correction factor (ICF) for oxygen, which accounts for the presence of ions above O$^{++}$. Our ICF uses the $(He^+ + He^{++})/He^+$ ratio with an exponent equal to 1, while a popular alternative, used by many groups, uses an exponent of $2/3$. The latter ICF would reduce the final oxygen abundance compared with the former. In Fig.~\ref{ocomparisons} the most egregious disparities occur for IC~418 and Cn~2-1, for which the SH10 abundance are much larger than ours. Both of these are low-excitation nebulae, so any ICF issue is moot. We do note that for two papers cited as abundance sources in SH10, the O/H values in those papers are much closer to ours: \citet{pbs06} derive O/H=$3.5 x10^{-4}$ for IC~418,\, bringing it down much closer to the unity line; and for Cn 2-1, \citet{chiappini09} calculate O/H= $4.7 x 10^{-4}$, which relocates it below and nearer to the unity line. 

As Fig.~\ref{ocomparisons} amply demonstrates, nebular abundances determined by different groups typically exhibit discrepancies beyond their nominal uncertainties. These discrepancies presumably arise from choices made during observation, calibration and/or data analysis, each of which contains multiple alternatives that could all contribute to divergent final outcomes. Several of us (RBCH, KBK and BB) are embarking on a collaboration with a number of other nebular abundance groups to thoroughly investigate the sources of disagreements in derived abundances. As discussed in \S4, these disparities currently represent a significant bottleneck hindering progress in using abundances to test ideas about galactic chemical evolution.

\pagebreak
\appendix

\clearpage

\begin{deluxetable}{lccccc}
\tablecolumns{3} \tablewidth{0pc} \tabletypesize{\scriptsize}
\tablenum{1} \tablecaption{Observing Log} \tablehead{
\colhead{Nebula Name} &\colhead{PNG} &\colhead{Date Observed} &
\colhead{Total Exp (sec)}& \colhead{Comments}}
 
\startdata

A 2&        122.1 -04.9     &Nov. 2007&    2700    &\\
A 8&        167.0 -00.9&    Nov. 2007&    3600    \\
A 12    &    204.0 -08.5&    Feb. 2007&    180    &\\
A 14    &    197.8 -03.3&    Feb. 2007&    1800&    PA=100$^{\circ}$&\\
BV5-2&    121.6 -00.0&    Oct. 2007&    3600&    PA=45$^{\circ}$\\
BV5-3&    131.4 -05.4&    Oct. 2007&    2700    &\\
H3-75&    193.6 -09.5&    Feb. 2007    &    1800&    \\
IC 289&    138.8 +02.8&    Feb. 2007&    180    &\\
IC 1747&    130.2 +01.3&    Jan. 2007&    480    &\\
IC 2149&    166.1 +10.4&    Jan. 2007&    90    &\\
K1-10&    229.6 -02.7&    Apr. 2007&    1800    &\\
K3-64&    151.4 +00.5&    Feb. 2007&    1860&    \\
K3-66&    167.4 -09.1&    Jan. 2007&    480    &\\
K3-67&    165.5 -06.5&    Jan. 2007    &360    &\\
K3-67&    165.5 -06.5&    Nov. 2007    &850    &\\
K3-70&    184.6 +00.6&    Jan. 2007    &1800&    \\
K3-90&    126.3 +02.9&    Oct. 2007    &2700&    \\
K3-91&    129.5 +04.5&    Nov. 2007    &1500&\\
K3-92&    130.4 +03.1&    Oct. 2007    &2700&    PA=45$^{\circ}$\\
K3-93&    132.4 +04.7&    Nov. 2007    &1800&\\
K3-94&    142.1 +03.4&    Feb. 2007    &960&\\
K4-47&    149.0 +04.4&    Nov. 2007    &1500    &PA=40$^{\circ}$\\
K4-48&    201.7 +02.5&    Jan. 2007    &720 &\\
M1-4    &    147.4 -02.3&    Jan. 2007    &540     &\\
M1-6    &    211.2 -03.5&    Jan. 2007    &600     &\\
M1-7    &    189.8 +07.7&    Jan. 2007    &600     &\\
M1-9&    212.0 +04.3&    Jan. 2007    &300     &\\
M1-11&    232.8 -04.7&    Feb. 2007    &190 &\\
M1-12&    235.3 -03.9&    Feb. 2007    &190     &\\
M1-14&    234.9 -01.4&    Apr. 2007    &600     &\\
M1-16&    226.7 +05.6&    Apr. 2007    &1200     &\\
M2-2    &    147.8 +04.1&    Jan. 2007&900        &\\
M4-18&    146.7 +07.6&    Jan. 2007    &420     &\\
NGC 1501&    144.5 +06.5&    Jan. 2007&    240 &\\
NGC 2346&    215.6 +03.6&    Feb. 2007    &180 &\\
PB1    &    226.4 -03.7&    Apr. 2007&    900 &\\
St3-1&    217.4 +02.0&    Feb. 2007&    240 &\\
YC2-5&    240.3 +07.0&    Apr. 2007    &600 &\\
\\ \hline \\
He2-15\tablenotemark{*}& 261.6 + 03.0& Nov. 2003&2100 B, 2400 R& CTIO&\\
K3-61\tablenotemark{*}& 096.3 + 02.3    & Aug. 2004&1200 B, 3000 R& KPNO&\\
M1-51\tablenotemark{*}& 020.9 - 01.1& Aug. 2004& 1800 B, 2100 R& KPNO&\\
NGC 6620\tablenotemark{*}& 005.9 - 06.2& Aug. 2004& 2160 B, 4180 R&CTIO&\\

\enddata
\tablenotetext{*}{The line strength and abundance data for these non-APO objects have been included as an addendum to the online tables.}
\end{deluxetable}

\clearpage

\begin{deluxetable}{lcr@{}lr@{}lr@{}lr@{}lr@{}lr@{}lr@{}lr@{}l}
\tabletypesize{\scriptsize}
\setlength{\tabcolsep}{0.07in}
\rotate
\tablecolumns{10}
\tablewidth{0in}
\tablenum{2}
\tablecaption{Fluxes and Intensities}
\tablehead{
\colhead{} & \colhead{} &
\multicolumn{4}{c}{A 2} &
\multicolumn{4}{c}{A 8} &
\multicolumn{4}{c}{A 12} &
\multicolumn{4}{c}{A 14} \\
\cline{3-6} \cline{7-10} \cline{11-14} \cline{15-18}  \\
\colhead{Line} &
\colhead{f($\lambda$)} &
\multicolumn{2}{c}{F($\lambda$)\tablenotemark{c}} &
\multicolumn{2}{c}{I($\lambda$)} &
\multicolumn{2}{c}{F($\lambda$)} &
\multicolumn{2}{c}{I($\lambda$)} &
\multicolumn{2}{c}{F($\lambda$)} &
\multicolumn{2}{c}{I($\lambda$)} &
\multicolumn{2}{c}{F($\lambda$)} &
\multicolumn{2}{c}{I($\lambda$)}}
\startdata
\[[O II] $\lambda$3727 & 0.292 & 79.2 & & 116 & $\pm$26 & 230 & & 416 & $\pm$93 & 273 & & 382 & $\pm$86 & 180 & & 326 & $\pm$73 \\
\[[Ne III] $\lambda$3869 & 0.252 & 76.6 & & 106 & $\pm$23 & 60.8 & & 102 & $\pm$21 & 82.3 & & 110 & $\pm$23 & 38.1 &:: & 63.5 & $\pm$33.86:: \\
He I + H8 $\lambda$3889 & 0.247 & 13.8 & & 19.0 & $\pm$4.00 & 12.5 & & 20.7 & $\pm$4.31 & 13.9 & & 18.4 & $\pm$3.86 & \nodata & & \nodata&  \\
\[[Ne III] $\lambda$3968 & 0.225 & 24.0 &\tablenotemark{a} & 32.0 & $\pm$9.80\tablenotemark{a} & 23.8 &\tablenotemark{a} & 37.5 & $\pm$10.59\tablenotemark{a} & 25.4 &\tablenotemark{a} & 32.9 & $\pm$9.90\tablenotemark{a} & \nodata & & \nodata&  \\
H$\epsilon$ $\lambda$3970 & 0.224 & 12.2 &\tablenotemark{a} & 16.3 &\tablenotemark{a} & 9.58 &\tablenotemark{a} & 15.1 &\tablenotemark{a} & 12.5 &\tablenotemark{a} & 16.2 &\tablenotemark{a} & \nodata & & \nodata&  \\
He I + He II $\lambda$4026 & 0.209 & 2.04 &:: & 2.68 & $\pm$1.42:: & \nodata & & \nodata&  & \nodata & & \nodata&  & \nodata & & \nodata&  \\
He II $\lambda$4100 & 0.188 & 0.452 &\tablenotemark{a} & 0.577 &\tablenotemark{a} & 0.144 &\tablenotemark{a} & 0.211 &\tablenotemark{a} & 0.249 &\tablenotemark{a} & 0.309 &\tablenotemark{a} & \nodata & & \nodata&  \\
H$\delta$ $\lambda$4101 & 0.188 & 20.3 &\tablenotemark{a} & 25.9 & $\pm$5.04\tablenotemark{a} & 25.1 &\tablenotemark{a} & 36.8 & $\pm$7.04\tablenotemark{a} & 19.5 &\tablenotemark{a} & 24.2 & $\pm$4.65\tablenotemark{a} & \nodata & & \nodata&  \\
He II $\lambda$4339 & 0.124 & 0.869 &\tablenotemark{a} & 1.02 &\tablenotemark{a} & 0.295 &\tablenotemark{a} & 0.380 &\tablenotemark{a} & 0.474 &\tablenotemark{a} & 0.546 &\tablenotemark{a} & 0.416 &\tablenotemark{a} & 0.536 &\tablenotemark{a} \\
H$\gamma$ $\lambda$4340 & 0.124 & 40.7 &\tablenotemark{a} & 47.8 & $\pm$8.36\tablenotemark{a} & 32.6 &\tablenotemark{a} & 42.0 & $\pm$7.25\tablenotemark{a} & 42.0 &\tablenotemark{a} & 48.5 & $\pm$8.41\tablenotemark{a} & 47.2 &\tablenotemark{a} & 60.7 & $\pm$10.50\tablenotemark{a} \\

\\
\vdots \\
\\

\[[S III] $\lambda$9069 & -0.670 & 92.2 & & 38.8 & $\pm$8.90 & 102 & & 26.2 & $\pm$5.96 & 38.9 & & 18.0 & $\pm$4.13 & 22.6 &: & 5.84 & $\pm$2.12: \\
P9 $\lambda$9228 & -0.610 & 9.70 & & 4.41 & $\pm$0.93 & \nodata & & \nodata&  & 4.78 & & 2.37 & $\pm$0.50 & \nodata & & \nodata&  \\
\[[S III] $\lambda$9532 & -0.632 & 188 & & 83.0 & $\pm$18.01 & 256 & & 71.1 & $\pm$15.30 & 108 & & 52.4 & $\pm$11.38 & 56.5 &: & 15.8 & $\pm$5.61: \\
P8 $\lambda$9546 & -0.633 & 8.35 & & 3.68 & $\pm$0.80 & \nodata & & \nodata&  & 6.47 & & 3.12 & $\pm$0.68 & \nodata & & \nodata&  \\
\[[C I] $\lambda$9824 & -0.653 & \nodata & & \nodata&  & \nodata & & \nodata&  & 11.9 &: & 5.62 & $\pm$2.03: & \nodata & & \nodata&  \\
\[[C I] $\lambda$9850 & -0.654 & \nodata & & \nodata&  & \nodata & & \nodata&  & 38.6 & & 18.2 & $\pm$4.08 & \nodata & & \nodata&  \\
c & & & 0.56 & & & &0.88& &  & &0.50& &  & &0.88 \\
H$\alpha$/H$\beta$ & & & 2.83 & & & &2.86 & & & &2.84 & & & &2.86 \\
log F$_{H\beta}$\tablenotemark{b} & & -13.59& &  & &-14.21& &  & &-12.42 & & & &-14.82 \\
\enddata
\tablenotetext{a}{Deblended.}
\tablenotetext{b}{ergs\ cm$^{-2}$ s$^{-1}$ in our extracted spectra}
\tablenotetext{c}{The F($\lambda$) values have been scaled to H$\beta$=100 using the observed F$_{H\beta}$ values. Intensities of strong lines have measurement uncertainties of $10\%$, single colons indicate uncertainties of $25\%$, double colons indicate uncertainties $50\%$, and triple colons indicate highly suspect measurements.}
\end{deluxetable}

\clearpage

\begin{deluxetable}{lcr@{}lcr@{}lcr@{}lcr@{}l}
\tabletypesize{\scriptsize}
\setlength{\tabcolsep}{0.07in}
\rotate
\tablecolumns{13}
\tablewidth{0in}
\tablenum{3}
\tablecaption{Ionic Abundances}
\tablehead{
\colhead{} &
\multicolumn{3}{c}{A 2} &
\multicolumn{3}{c}{A 8} &
\multicolumn{3}{c}{A 12} &
\multicolumn{3}{c}{A 14} \\
\cline{2-4} \cline{5-7} \cline{8-10} \cline{11-13}  \\
\colhead{Ion} &
\colhead{T$_{\mathrm{used}}$\tablenotemark{a}} &
\multicolumn{2}{c}{Abundance\tablenotemark{b}} &
\colhead{T$_{\mathrm{used}}$} &
\multicolumn{2}{c}{Abundance} &
\colhead{T$_{\mathrm{used}}$} &
\multicolumn{2}{c}{Abundance} &
\colhead{T$_{\mathrm{used}}$} &
\multicolumn{2}{c}{Abundance}
}
\startdata
He$^{+}$ & [O III] & 8.59 & $\pm$1.08(-2) & [O III] & 7.23 & $\pm$3.66(-2) & [O III] & 0.109 & $\pm$0.014 & [O III] & 0.172 & $\pm$0.087\\
He$^{+2}$ & [O III] & 4.00 & $\pm$1.31(-2) & [O III] & 1.88 & $\pm$0.28(-2) & [O III] & 2.14 & $\pm$0.32(-2) & [O III] & 2.40 & $\pm$1.25(-2)\\
icf(He) &  & 1.00 & &  & 1.00 & &  & 1.00 & &  & 1.00 &\\
\\
O$^{0}$(6300) & [N II] & \nodata &  & [N II] & \nodata &  & [N II] & $^{*}$2.58 & $\pm$0.67(-5) & [N II] & $^{*}$2.96 & $\pm$1.24(-5)\\
O$^{0}$(6363) & [N II] & \nodata &  & [N II] & \nodata &  & [N II] & $^{*}$2.67 & $\pm$0.70(-5) & [N II] & $^{*}$4.08 & $\pm$1.71(-5)\\
O$^{0}$ & wm & \nodata &  & wm & \nodata &  & wm & 2.60 & $\pm$0.66(-5) & wm & 3.31 & $\pm$1.37(-5)\\
O$^{+}$(3727) & [N II] & $^{*}$3.43 & $\pm$0.75(-5) & [N II] & $^{*}$9.92 & $\pm$7.11(-5) & [N II] & $^{*}$7.40 & $\pm$1.89(-5) & [N II] & $^{*}$5.35 & $\pm$2.44(-5)\\
O$^{+}$(7325) & [N II] & $^{*}$2.92 & $\pm$1.63(-5) & [N II] & $^{*}$2.70 & $\pm$3.30(-5) & [N II] & $^{*}$6.21 & $\pm$2.75(-5) & [N II] & $^{*}$4.29 & $\pm$3.12(-5)\\
O$^{+}$ & wm & 3.41 & $\pm$0.70(-5) & wm & 9.70 & $\pm$6.98(-5) & wm & 7.32 & $\pm$1.83(-5) & wm & 5.24 & $\pm$2.44(-5)\\
O$^{+2}$(5007) & [O III] & $^{*}$2.62 & $\pm$0.66(-4) & [O III] & $^{*}$2.95 & $\pm$0.40(-4) & [O III] & $^{*}$2.22 & $\pm$0.56(-4) & [O III] & $^{*}$1.72 & $\pm$0.23(-4)\\
O$^{+2}$(4959) & [O III] & $^{*}$2.50 & $\pm$0.51(-4) & [O III] & $^{*}$2.79 & $\pm$0.38(-4) & [O III] & $^{*}$2.02 & $\pm$0.40(-4) & [O III] & $^{*}$1.65 & $\pm$0.23(-4)\\
O$^{+2}$(4363) & [O III] & $^{*}$2.62 & $\pm$0.66(-4) & [O III] & \nodata &  & [O III] & $^{*}$2.22 & $\pm$0.56(-4) & [O III] & \nodata & \\
O$^{+2}$ & wm & 2.59 & $\pm$0.61(-4) & wm & 2.91 & $\pm$0.35(-4) & wm & 2.18 & $\pm$0.51(-4) & wm & 1.70 & $\pm$0.21(-4)\\
icf(O) &  & 1.47 & $\pm$0.16 &  & 1.26 & $\pm$0.14 &  & 1.20 & $\pm$0.04 &  & 1.14 & $\pm$0.10\\
\\
\vdots \\
\\
S$^{+}$ & [N II] & $^{*}$5.31 & $\pm$0.84(-7) & [N II] & $^{*}$9.23 & $\pm$3.54(-7) & [N II] & $^{*}$8.85 & $\pm$1.89(-7) & [N II] & $^{*}$1.77 & $\pm$0.46(-6)\\
S$^{+}$(6716) & [N II] & 5.31 & $\pm$0.85(-7) & [N II] & 9.25 & $\pm$3.53(-7) & [N II] & 8.85 & $\pm$1.89(-7) & [N II] & 1.77 & $\pm$0.46(-6)\\
S$^{+}$(6731) & [N II] & 5.31 & $\pm$0.84(-7) & [N II] & 9.21 & $\pm$3.56(-7) & [N II] & 8.85 & $\pm$1.89(-7) & [N II] & 1.76 & $\pm$0.47(-6)\\
S$^{+2}$(9069) & [S III] & $^{*}$4.23 & $\pm$1.55(-6) & [O III] & $^{*}$4.01 & $\pm$0.86(-6) & [S III] & $^{*}$2.10 & $\pm$0.65(-6) & [O III] & $^{*}$8.92 & $\pm$3.17(-7)\\
S$^{+2}$(6312) & [S III] & $^{*}$4.23 & $\pm$1.55(-6) & [O III] &  $^{*}$\nodata &  & [S III] & $^{*}$2.10 & $\pm$0.65(-6) & [O III] &  $^{*}$\nodata & \\
S$^{+2}$ & wm & 4.23 & $\pm$1.55(-6) & wm & \nodata &  & wm & 2.10 & $\pm$0.65(-6) & wm & \nodata & \\
icf(S) &  & 1.37 & $\pm$0.06 &  & 1.19 & $\pm$0.08 &  & 1.19 & $\pm$0.03 &  & 1.19 & $\pm$0.05\\
\\
\enddata
\tablenotetext{a} {\ wm indicates the calculated mean value for the ionic abundance weighted by the observed flux}
\tablenotetext{b}{\ * indicates those values included in the weighted mean}
\end{deluxetable}

\clearpage

\begin{deluxetable}{lr@{}llr@{}llr@{}llr@{}ll}
\tabletypesize{\small}
\setlength{\tabcolsep}{0.07in}
\rotate
\tablecolumns{13}
\tablewidth{0in}
\tablenum{4}
\tablecaption{Temperatures and Densities}
\tablehead{
\colhead{} &
\multicolumn{3}{c}{A 2} &
\multicolumn{3}{c}{A 8} &
\multicolumn{3}{c}{A 12} &
\multicolumn{3}{c}{A 14} \\
\cline{2-4} \cline{5-7} \cline{8-10} \cline{11-13}  \\
\colhead{Parameter} &
\multicolumn{2}{c}{Value\tablenotemark{a}} &
\colhead{Notes\tablenotemark{b}} &
\multicolumn{2}{c}{Value} &
\colhead{Notes} &
\multicolumn{2}{c}{Value} &
\colhead{Notes} &
\multicolumn{2}{c}{Value} &
\colhead{Notes}
}
\startdata
T$_{[O~III]}$ & 11710 & $\pm$649 &  & 10000 & & Default. & 11430 & $\pm$618 &  & 10000 & & Default.\\
T$_{[N~II]}$ & 10300 & & Default. & 11580 & $\pm$2791 &  & 12160 & $\pm$932 &  & 13270 & $\pm$2306 & \\
T$_{[O~II]}$ & 9453 & $\pm$3083 &  & 6425 & $\pm$1409 &  & 10850 & $\pm$2680 &  & 11450 & $\pm$2833 & \\
T$_{[S~III]}$ & 11050 & $\pm$1923 & Used 9069.  & \nodata &  &  & 12460 & $\pm$1275 & Used 9532.  & \nodata &  & \\
N$_{e, [S~II]}$ & 73.2 & $\pm$179.30 &  & 75.2 & $\pm$169.90 &  & 301 & $\pm$244 &  & 55.1 & $\pm$164.50 & \\
\enddata

\tablenotetext{a}{Temperatures and densities given in kelvin and cm$^{-3}$ respectively.}
\tablenotetext{b}{If [N~II] $\lambda$5755 is unavailable, ${T_{[N~II]}}$ is estimated from ${T_{[O~III]}}$ (Kwitter \& Henry 2001).  Since telluric absorption tends to affect only one or the other of [S~III] $\lambda$9069 and $\lambda$9532, if the observed flux ratio $\lambda$9532/$\lambda$9069 $\geq$ 2.48 (the theoretical value) then $\lambda$9532 is used in the ${T_{[S~III]}}$ calculation; otherwise $\lambda$9069 is used. Default ${T_{[N~II]}}$ and ${N_{e [Cl~III]}}$ values and the high density limit for ${N_{e [S~II]}}$ are based on criteria discussed in Kwitter \& Henry (2001).} 

\end{deluxetable}

\clearpage

\begin{deluxetable}{lr@{}lr@{}lr@{}lr@{}lr@{}lr@{}l}
\tabletypesize{\small}
\setlength{\tabcolsep}{0.07in}
\tablecolumns{11}
\tablewidth{0in}
\tablenum{5}
\tablecaption{Total Elemental Abundances}
\tablehead{
\colhead{Parameter}\tablenotemark{a} &
\multicolumn{2}{c}{A 2} &
\multicolumn{2}{c}{A 8} &
\multicolumn{2}{c}{A 12} &
\multicolumn{2}{c}{A 14} &
\multicolumn{2}{c}{Solar Ref}\tablenotemark{b} &
\multicolumn{2}{c}{Orion Ref}\tablenotemark{c}
}
\startdata
He/H & 0.126 & $\pm$0.018 & 9.11 & $\pm$3.68(-2) & 0.130 & $\pm$0.015 & 0.196 & $\pm$0.088 & 8.51(-2) & & 9.73(-2) &\\
N/H & 1.34 & $\pm$0.47(-4) & 1.48 & $\pm$0.40(-4) & 1.32 & $\pm$0.37(-4) & 1.05 & $\pm$0.24(-3) & 6.76(-5) & & 5.37(-5) &\\
N/O & 0.313 & $\pm$0.088 & 0.302 & $\pm$0.126 & 0.380 & $\pm$0.090 & 4.12 & $\pm$1.27 & 0.138 & & 0.100 &\\
O/H & 4.29 & $\pm$1.05(-4) & 4.89 & $\pm$1.19(-4) & 3.48 & $\pm$0.71(-4) & 2.54 & $\pm$0.47(-4) & 4.90(-4) & & 5.37(-4) &\\
Ne/H & 9.93 & $\pm$2.71(-5) & 1.76 & $\pm$0.56(-4) & 1.08 & $\pm$0.25(-4) & 9.79 & $\pm$5.48(-5) & 8.51(-5) & & 1.12(-4)) &\\
Ne/O & 0.231 & $\pm$0.037 & 0.361 & $\pm$0.063 & 0.311 & $\pm$0.051 & 0.385 & $\pm$0.201 & 0.174 & & 0.209 &\\
S/H & 6.52 & $\pm$2.37(-6) & 5.88 & $\pm$1.18(-6) & 3.54 & $\pm$0.97(-6) & 3.15 & $\pm$0.69(-6) & 1.32(-5) & & 1.66(-5) &\\
S/O & 1.52 & $\pm$0.61(-2) & 1.20 & $\pm$0.41(-2) & 1.02 & $\pm$0.31(-2) & 1.24 & $\pm$0.33(-2) & 2.69(-2) & & 3.09(-2) &\\
Cl/H & \nodata &  & \nodata &  & 4.14 & $\pm$1.58(-8) & \nodata &  & 3.16(-7) & & 2.88(-7) &\\
Cl/O & \nodata &  & \nodata &  & 1.19 & $\pm$0.48(-4) & \nodata &  & 6.46(-4) & & 5.37(-4) &\\
Ar/H & 1.98 & $\pm$0.43(-6) & 2.66 & $\pm$0.71(-6) & 1.30 & $\pm$0.26(-6) & 1.46 & $\pm$0.28(-6) & 2.51(-6) & & 4.17(-6) &\\
Ar/O & 4.61 & $\pm$0.96(-3) & 5.44 & $\pm$1.47(-3) & 3.72 & $\pm$0.78(-3) & 5.75 & $\pm$1.22(-3) & 5.13(-3) & & 7.76(-3) &\\
\enddata

\tablenotetext{a}{all abundances are in linear form} 
\tablenotetext{b}{Solar values taken from Asplund et al. 2009.} 
\tablenotetext{c}{Orion values taken from \citet{esteban04}} 

\end{deluxetable}

\begin{deluxetable}{lccccc}
\tablecolumns{6}
\tablecaption{Data Used In Analysis\label{analdata}}
 \tabletypesize{\normalsize}
 \setlength{\tabcolsep}{0.07in}
 \tablewidth{0in}
 \tablenum{6}
\tablehead{
\colhead{PN Name} &
 \colhead{D\tablenotemark{a}} & 
 \colhead{R\tablenotemark{b}} & 
 \colhead{D$_{SSV}$\tablenotemark{c}} & 
 \colhead{R$_{SSV}$\tablenotemark{b}} & 
 \colhead{12+log(O/H)}
 }
\startdata
A12	&	2.04	&	10.44	&	2.06	&	10.46	&	8.54	\\
Cn2-1	&	3.86	&	4.67	&	8.00	&	0.76	&	8.80	\\
Cn3-1	&	3.58	&	6.14	&	6.04	&	5.31	&	8.60	\\
Fg1	&	3.1	&	7.96	&	3.13	&	7.96	&	8.51	\\
H2-18	&	3.90	&	4.65	&	\nodata	&	\nodata	&	8.75	\\
H3-75	&	2.70	&	11.14	&	\nodata	&	\nodata	&	8.58	\\
Hb4	&	2.08	&	6.43	&	5.04	&	3.48	&	8.82	\\
Hb6	&	1.66	&	6.86	&	\nodata	&	\nodata	&	8.66	\\
He2-111	&	3.48	&	6.52	&	3.51	&	6.51	&	8.44	\\
He2-115	&	1.95	&	7.09	&	\nodata	&	\nodata	&	8.50	\\
He2-141	&	3.44	&	6.00	&	3.47	&	5.98	&	8.80	\\
He2-15	&	2.14	&	9.06	&	2.17	&	9.07	&	8.37	\\
He2-158	&	19.74	&	13.23	&	19.90	&	13.38	&	8.43	\\
He2-21	&	7.3	&	10.67	&	12.50	&	14.42	&	8.38	\\
He2-37	&	2.98	&	8.78	&	3.00	&	8.78	&	8.93	\\
He2-48	&	5.10	&	8.88	&	\nodata	&	\nodata	&	8.53	\\
He2-55	&	2.60	&	8.16	&	\nodata	&	\nodata	&	8.76	\\
Hu1-2	&	1.48	&	8.54	&	\nodata	&	\nodata	&	8.23	\\
Hu2-1	&	2.5	&	7.22	&	\nodata	&	\nodata	&	8.43	\\
IC1297	&	3.76	&	5.01	&	4.91	&	3.94	&	8.80	\\
IC1747	&	2.94	&	10.64	&	2.99	&	10.68	&	8.57	\\
IC2149	&	1.59	&	10.03	&	3.26	&	11.64	&	8.35	\\
IC2165	&	2.21	&	10.22	&	3.65	&	11.42	&	8.40	\\
IC2501	&	2.62	&	8.40	&	\nodata	&	\nodata	&	8.57	\\
IC2621	&	1.87	&	8.00	&	4.87	&	8.09	&	8.61	\\
IC3568	&	2.71	&	9.91	&	2.74	&	9.93	&	8.51	\\
IC418	&	0.61	&	8.96	&	\nodata	&	\nodata	&	8.28	\\
IC4593	&	3.19	&	6.40	&	3.23	&	6.38	&	8.62	\\
IC4776	&	3.9	&	4.71	&	4.97	&	3.67	&	8.61	\\
IC5217	&	4.65	&	10.40	&	5.37	&	10.84	&	8.51	\\
J320	&	6.06	&	14.22	&	6.12	&	14.27	&	8.32	\\
J900	&	2.76	&	11.19	&	4.90	&	13.30	&	8.47	\\
K3-61	&	7.33	&	11.81	&	7.41	&	11.87	&	8.60	\\
K3-64	&	9.82	&	17.75	&	\nodata	&	\nodata	&	8.42	\\
K3-67	&	3.4	&	11.80	&	3.43	&	11.83	&	8.12	\\
K3-70	&	12.12	&	20.60	&	19.40	&	27.88	&	8.08	\\
K3-91	&	8.44	&	15.30	&	8.52	&	15.37	&	8.34	\\
K3-92	&	6.81	&	13.91	&	6.87	&	13.96	&	8.50	\\
K3-93	&	7.45	&	14.58	&	\nodata	&	\nodata	&	8.40	\\
K3-94	&	6.51	&	14.20	&	6.57	&	14.26	&	8.51	\\
K4-47	&	8.53	&	16.39	&	8.61	&	16.46	&	7.69	\\
K4-48	&	7.77	&	15.97	&	13.50	&	21.61	&	8.51	\\
M1-13	&	5.32	&	12.48	&	5.38	&	12.53	&	8.70	\\
M1-14	&	3.95	&	11.24	&	6.42	&	13.27	&	8.46	\\
M1-16	&	5.45	&	12.84	&	9.77	&	16.74	&	8.52	\\
M1-17	&	7.36	&	14.42	&	10.80	&	17.56	&	8.69	\\
M1-34	&	4.99	&	3.54	&	5.05	&	3.48	&	8.70	\\
M1-4	&	2.99	&	11.13	&	6.60	&	14.50	&	8.40	\\
M1-40	&	1.81	&	6.71	&	\nodata	&	\nodata	&	8.75	\\
M1-42	&	5.46	&	3.08	&	5.51	&	3.03	&	8.55	\\
M1-5	&	2.92	&	11.41	&	\nodata	&	\nodata	&	8.16	\\
M1-50	&	4.36	&	4.43	&	5.98	&	3.12	&	8.73	\\
M1-51	&	1.75	&	6.89	&	2.29	&	6.41	&	8.94	\\
M1-54	&	3.78	&	4.99	&	3.81	&	4.96	&	8.64	\\
M1-7	&	5.91	&	14.31	&	5.96	&	14.36	&	8.67	\\
M1-74	&	4.12	&	6.81	&	\nodata	&	\nodata	&	8.62	\\
M1-8	&	3.39	&	11.55	&	3.42	&	11.58	&	8.65	\\
M1-80	&	5.5	&	11.44	&	5.55	&	11.47	&	8.87	\\
M1-9	&	4.88	&	12.89	&	11.60	&	19.31	&	8.28	\\
M2-52	&	4.41	&	10.46	&	4.45	&	10.49	&	8.33	\\
M2-55	&	2.21	&	9.67	&	2.23	&	9.68	&	8.93	\\
M3-15	&	0.29	&	4.47	&	\nodata	&	\nodata	&	8.87	\\
M3-2	&	8.82	&	14.91	&	8.93	&	15.01	&	7.83	\\
M3-3	&	5.75	&	13.33	&	5.81	&	13.38	&	8.47	\\
M3-4	&	6.33	&	12.82	&	6.39	&	12.87	&	8.58	\\
M3-5	&	7.14	&	13.18	&	7.21	&	13.24	&	8.41	\\
M3-6	&	3.29	&	9.92	&	4.22	&	10.47	&	8.70	\\
Me1-1	&	4.62	&	6.77	&	6.76	&	6.93	&	8.69	\\
Me2-2	&	5	&	10.55	&	\nodata	&	É	&	8.41	\\
Mz2	&	2.34	&	6.60	&	2.36	&	6.58	&	8.95	\\
Mz3	&	1.27	&	7.41	&	1.45	&	7.26	&	8.16	\\
NGC 2346	&	1.36	&	9.64	&	1.37	&	9.64	&	8.60	\\
NGC 2371	&	1.54	&	9.93	&	1.55	&	9.94	&	8.79	\\
NGC 2392	&	1.25	&	9.64	&	1.26	&	9.65	&	8.39	\\
NGC 2438	&	1.2	&	9.29	&	1.22	&	9.30	&	8.61	\\
NGC 2440	&	1.35	&	9.34	&	1.36	&	9.35	&	8.62	\\
NGC 2452	&	2.81	&	10.08&	2.84	&	10.10&	8.87	\\
NGC 2792	&	3.02	&	9.23	&	3.05	&	9.24	&	8.82	\\
NGC 2867	&	1.84	&	8.44	&	2.23	&	8.48	&	8.64	\\
NGC 3195	&	1.96	&	7.85	&	1.98	&	7.85	&	8.71	\\
NGC 3211	&	2.87	&	8.17	&	2.90	&	8.17	&	8.84	\\
NGC 3242	&	1.08	&	8.69	&	1.09	&	8.69	&	8.57	\\
NGC 3587	&	0.62	&	8.79	&	0.62	&	8.79	&	8.54	\\
NGC 3918	&	1.01	&	8.13	&	1.64	&	7.96	&	8.67	\\
NGC 5307	&	3.2	&	6.79	&	3.24	&	6.78	&	8.50	\\
NGC 5315	&	1.24	&	7.78	&	\nodata	&	\nodata	&	8.64	\\
NGC 5882	&	1.68	&	7.15	&	2.36	&	6.65	&	8.68	\\
NGC 6210	&	2.03	&	7.41	&	2.28	&	7.29	&	8.68	\\
NGC 6302	&	0.53	&	7.98	&	0.74	&	7.77	&	8.21	\\
NGC 6309	&	2.53	&	6.10	&	2.74	&	5.90	&	8.77	\\
NGC 6369	&	0.66	&	7.84	&	1.09	&	7.42	&	8.71	\\
NGC 6445	&	1.37	&	7.15	&	1.38	&	7.14	&	8.83	\\
NGC 6439	&	4.12	&	4.54	&	6.33	&	2.61	&	8.89	\\
NGC 650	   	&	0.74	&	8.99	&	7.46	&	14.41&	8.80	\\
NGC 6537	&	0.9	&	7.62	&	\nodata	&	\nodata	&	8.32	\\
NGC 6563	&	1.63	&	6.88	&	1.65	&	6.86	&	8.66	\\
NGC 6565	&	4.62	&	3.91	&	4.66	&	3.87	&	8.75	\\
NGC 6567	&	2.37	&	6.20	&	3.61	&	5.02	&	8.36	\\
NGC 6572	&	0.71	&	7.94	&	1.74	&	7.16	&	8.58	\\
NGC 6578	&	2.31	&	6.25	&	3.64	&	4.97	&	8.83	\\
NGC 6620	&	8.75	&	0.89	&	8.84	&	0.92	&	8.86	\\
NGC 6629	&	1.95	&	8.50	&	2.37	&	6.18	&	8.78	\\
NGC 6720	&	0.87	&	6.89	&	0.88	&	8.15	&	8.79	\\
NGC 6741	&	2.05	&	6.40	&	3.73	&	5.79	&	8.61	\\
NGC 6751	&	2.56	&	6.20	&	2.89	&	6.15	&	8.63	\\
NGC 6790	&	1.54	&	7.35	&	\nodata	&	\nodata	&	8.60	\\
NGC 6803	&	2.99	&	6.80	&	5.27	&	6.19	&	8.74	\\
NGC 6826	&	1.58	&	8.47	&	1.59	&	8.47	&	8.57	\\
NGC 6853	&	0.26	&	8.50	&	0.26	&	8.38	&	8.72	\\
NGC 6881	&	2.47	&	8.19	&	5.34	&	8.75	&	8.65	\\
NGC 6884	&	2.11	&	8.47	&	3.83	&	8.82	&	8.69	\\
NGC 6886	&	3.1	&	7.46	&	4.35	&	7.37	&	8.65	\\
NGC 6891	&	3.19	&	7.13	&	3.61	&	7.04	&	8.57	\\
NGC 7008	&	0.86	&	8.59	&	0.87	&	8.59	&	8.76	\\
NGC 7009	&	1.2	&	7.74	&	1.33	&	7.66	&	8.72	\\
NGC 7026	&	1.9	&	8.68	&	2.35	&	8.78	&	8.81	\\
NGC 7027	&	0.27	&	8.48	&	\nodata	&	\nodata	&	8.52	\\
NGC 7293	&	0.16	&	8.40	&	0.16	&	8.43	&	8.66	\\
NGC 7354	&	1.27	&	8.97	&	1.70	&	9.16	&	8.71	\\
PB6	&	4.38	&	8.94	&	4.42	&	8.95	&	8.69	\\
PC14	&	5.74	&	4.01	&	5.80	&	3.98	&	8.84	\\
Pe1-18	&	3.56	&	5.58	&	\nodata	&	\nodata	&	8.57	\\
St3-1	&	5.80	&	13.57	&	\nodata	&	\nodata	&	8.58	\\
Th2-A	&	2.45	&	7.32	&	2.47	&	7.31	&	8.71	
\enddata
\tablenotetext{a}{Most of the heliocentric distances in this column were taken directly from \citet{cahn92}. The sources for the remaining objects are provided in Section 4.}
\tablenotetext{b}{Galactocentric distances were computed from the heliocentric distances in column 2 using the relation: $R=\{R_{\odot}^2 + [cos(b) \times D]^2 - 2 \times R_{\odot} \times D \times cos(l) \times cos(b)\}^{1/2}$, where $R$ and $D$ are galactocentric and heliocentric distances, respectively, and $b$ and $l$ are galactic latitude and longitude, respectively. We assumed that $R_{\odot}=8.5$~kpc.}
\tablenotetext{c}{Heliocentric distances in this column were taken directly from \citet{stanghellini08}.}
\end{deluxetable}
\begin{deluxetable}{lccc}
\tablecolumns{4}
\tablecaption{Trial Least Squares Fits\label{lsf}}
 \tabletypesize{\normalsize}
 \setlength{\tabcolsep}{0.07in}
 \tablewidth{0in}
 \tablenum{7}
\tablehead{
\colhead{f(O)} &
 \colhead{f($R_g$)} & 
 \colhead{b} & 
 \colhead{$q_{\chi^2}$}
 }
\startdata
         1    &     0 &   -0.041 & 1.95E-28 \\
         1    &   0.1 &   -0.051&  4.70E-16 \\
      1.25    &     0 &   -0.041 & 3.77E-10 \\
      1.25   &    0.1  &   -0.049&  4.92E-06 \\
         1    &   0.2  &  -0.066 &  0.00071 \\
       1.5   &      0 &   -0.041 &   0.0030 \\
       1.5   &    0.1 &   -0.047 &    0.050 \\
      1.25   &    0.2 &   -0.061 &     0.13 \\
      1.75    &     0 &   -0.041 &     0.42 \\
      1.75   &    0.1 &   -0.045 &     0.71 \\
       1.5   &    0.2 &   -0.056  &     0.75 \\
         1   &    0.3 &   -0.074 &     0.78 \\
         2    &     0 &    -0.041  &    0.96 \\
      1.25  &     0.3 &   -0.070 &     0.98 \\
         2   &    0.1 &   -0.045 &     0.99 \\
      1.75  &     0.2 &   -0.053 &     0.99 \\
         2   &    0.2 &   -0.051 &        1 \\
       1.5  &     0.3 &   -0.066  &       1 \\
         2    &   0.3 &   -0.059 &        1 \\
      1.75   &    0.3 &   -0.062 &        1
\enddata
\end{deluxetable}
\begin{deluxetable}{lcccccccc}
\tabletypesize{\small}
\setlength{\tabcolsep}{0.07in}
\tablecolumns{9}
\tablewidth{0in}
\tablenum{8}
\tablecaption{Oxygen Gradients\label{gradients}}
\tablehead{
\colhead{Sample\tablenotemark{a}} &
\colhead{n} &
\colhead{a} &
\colhead{b} &
\colhead{$\chi^2$} &
\colhead{$\chi^2_{\nu}$} &
\colhead{q$_{\chi^2}$} &
\colhead{r} &
\colhead{p$_r$}
}
\startdata
Total & 124 & 9.09$\pm$.05 & -0.058$\pm$.006 & 121.8 & 1.00 & 0.49 & -0.54 & 3.3(-11) \\
Type I & 48 &9.10$\pm$.07 &-0.061$\pm$.008 &62.7 &1.36 &0.051 &-0.59 &3.5(-6) \\
Type II & 75 &9.05$\pm$.08 &-0.053$\pm$.010 &58.6 &0.80 &0.89 &-0.44 &4.8(-5) \\
R$<10$kpc &85  &9.04$\pm$.09 &-0.054$\pm$.013 &73.3 &0.88 &0.77 &-0.21 &4.8(-2) \\
R$>10$kpc &38  &9.92$\pm$10.10 &-0.12$\pm$.14 &24.2 &0.67 &0.93 &-0.39 &1.2(-2) \\
H II Regions & 73 &9.24$\pm$.07 &-0.078$\pm$.007 &63.5 &0.89 &0.72 &-0.80 &1.2(-20) \\
B V Stars &24 &9.63$\pm$.11 &-0.080$\pm$.01 &33.3 &1.51 &0.058 &-0.64 &2.6(-4) 
\enddata
\tablenotetext{a}{The first five groups refer to the PN sample presented here, where the abundance uncertainty for each object was increased by 40\% and the error in the galactocentric distance was assumed to be 20\%. The H~II region sample was assembled from studies by \citet{afflerbach97}, \citet{vilchez96}, and \citet{deharveng00}, and the B V star sample was compiled from papers by \citet{smartt01} and \citet{rolleston00}. }
\end{deluxetable}
\begin{deluxetable}{lcccc}
\tabletypesize{\small}
\setlength{\tabcolsep}{0.07in}
\tablecolumns{5}
\tablewidth{0in}
\tablenum{9}
\tablecaption{Oxygen Gradient Comparison\label{gradientscom}}
\tablehead{
\colhead{Source\tablenotemark{a}} &
\colhead{N} &
\colhead{$\Delta$log(O/H)/$\Delta$R (dex/kpc)} &
\colhead{ Object Type} &
\colhead{$\Delta R_g$ (kpc)} 
}
\startdata
This Paper & 124 & -0.058$\pm$.006 & PN & 0.9-21 \\
S10 & 145 & -0.023$\pm$.006 & PN &3-21\\
H04 &79 &-0.037$\pm$.008 & PN &2-17\\
C04 &80 &-0.05 $\pm$\nodata & PN & 4-14\\
P06 &83 &-0.016$\pm$.008 & PN &0.5-15 \\
R06 & 70 & -0.060$\pm$.010 & H II (optical) &5-18 \\
R06 & 68 & -0.041$\pm$.014 & H II (FIR) &0.1-15\\
P09 & 265 & -0.051$\pm$.004\tablenotemark{b} & Cepheids &5-17 \\
SR97 & 47 & -0.07$\pm$.01 & B V Stars &6-18 

\enddata

\tablenotetext{a}{S10: \citet{stanghellini10}; H04: \citet{hkb04}; C04: \citet{costa04}; P06: \citet{perinotto06}; R06: \citet{rudolph06}; P09: \citet{pedicelli09}; SR97: \citet{smartt97}}
\tablenotetext{b}{$\Delta$log(Fe/H)/$\Delta$R (dex/kpc)}

\end{deluxetable}

\clearpage

\end{document}